\documentclass{article}
\usepackage{amsmath,amssymb}
\usepackage{amsfonts}
\usepackage{graphicx}
\usepackage{geometry}
\usepackage{xcolor}
\usepackage[colorlinks=true]{hyperref}
\usepackage{ulem}
\usepackage{float}
\usepackage{backref}
\usepackage[numbers]{natbib}

\newcommand{\be}{\begin{equation}}
\newcommand{\ee}{\end{equation}}

\newcommand{\K}{\mathrm{K}}
\newcommand{\E}{\mathrm{E}}
\newcommand{\hb}{{\overline{h}}}
\newcommand{\ub}{{\overline{u}}}
\newcommand{\kt}{{\widetilde{k}}}
\newcommand{\omegat}{{\widetilde{\omega}}}

\title{Solitary wave-mean flow interaction in strongly nonlinear dispersive shallow water waves}
\author{Thibault CONGY\thanks{Department of Mathematics, Physics and Electrical Engineering, Northumbria University, Newcastle upon Tyne, NE1 8ST, UK, thibault.congy@northumbria.ac.uk}, Gennady EL\thanks{Department of Mathematics, Physics and Electrical Engineering, Northumbria University, Newcastle upon Tyne, NE1 8ST, UK, gennady.el@northumbria.ac.uk}, Sergey GAVRILYUK\thanks{Aix Marseille Univ, CNRS, IUSTI,   UMR 7343,  Marseille, France,   sergey.gavrilyuk@univ-amu.fr},  Mark HOEFER\thanks{Department of Applied Mathematics, University of Colorado, Boulder, Colorado, USA, hoefer@colorado.edu},\\ 
Keh-Ming SHYUE\thanks{Ocean Center, National Taiwan University,
                Taipei 106, Taiwan, shyue@ntu.edu.tw}}

\begin{document}
\maketitle
\begin{abstract}
  The interaction of a solitary wave and a slowly varying mean
  background or flow for the Serre-Green-Naghdi (SGN) equations is
  studied using Whitham modulation theory.  The exact form of the
  three SGN-Whitham modulation equations---two for the mean horizontal
  velocity and depth decoupled from one for the solitary wave
  amplitude field---are obtained exactly in the solitary wave limit.
  Although the three equations are not diagonalizable, the restriction
  of the full system to simple waves for the mean equations is
  diagonalized in terms of Riemann invariants.  The Riemann invariants
  are used to analytically describe the head-on and overtaking
  interactions of a solitary wave with a rarefaction wave and dispersive shock wave (DSW), leading to
  scenarios of solitary wave trapping or transmission by the mean
  flow.  The analytical results for overtaking interactions prove that
  a simpler, approximate approach based on the DSW
  fitting method is accurate to the second order in solitary wave
  amplitude, beyond the first order accurate Korteweg-de Vries
  approximation.  The analytical results also accurately predict the SGN DSW's solitary wave edge amplitude and speed.  The analytical results are favourably compared with
  corresponding numerical solutions of the full SGN equations.
  Because the SGN equations model the bi-directional propagation of
  strongly nonlinear, long gravity waves over a flat bottom, the
  analysis presented here describes large amplitude solitary wave-mean
  flow interactions in shallow water waves.
\end{abstract}

\section{Introduction} 

\begin{figure}
  \centering
  \includegraphics[width=12cm]{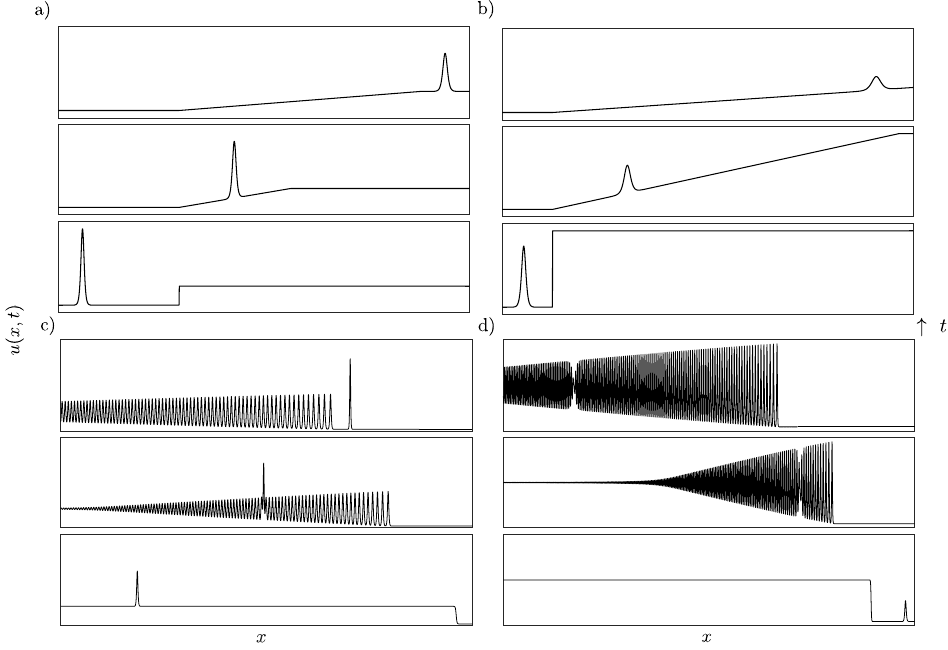}
    \caption{Scenarios of the solitary wave-mean flow interaction in shallow water waves. a), c): solitary wave transmission; b), d) solitary wave trapping. Reproduced, with permission, from \cite{ablowitz_solitonmean_2023}.}  
  \label{fig:solitary-wave-mean-schematic}
\end{figure}

A fundamental and important problem in continuum mechanics is the
interaction of waves with the inhomogeneity of the medium through
which they propagate.  The waves could be linear or nonlinear and, in
many cases, the inhomogeneity can be modelled as externally imposed
through prescribed variable coefficients to a wave-type partial
differential equation (PDE) \cite{chew_waves_1999}.  When the waves
and the medium are dynamically coupled, a common occurrence in
geophysical fluid dynamics \cite{buhler_waves_2014}, the problem
becomes more challenging to describe analytically.  A natural
framework to approach this class of problems is to utilize scale
separation and derive equations separating the motion of waves and
averaged quantities, e.g., the mean fluid density and velocity.  While
the focus of analytical studies was historically on linear or weakly
nonlinear waves interacting with mean flows \cite{buhler_waves_2014},
a recent body of work has emerged for the case where the waves are
strongly nonlinear, i.e., solitons or solitary waves; see, e.g., the
review \cite{ablowitz_solitonmean_2023}.  A schematic of such 
wave-mean flow scenarios involving solitary wave interaction with rarefaction and dispersive shock waves is shown in
Fig.~\ref{fig:solitary-wave-mean-schematic}.

For the asymptotic analysis of solitary wave-mean flow interaction,
the mean flow is assumed to vary on a much slower spatial scale than
the solitary wave width.  In this case, the mean flow evolution
decouples from solitary wave motion so that the interaction is
one-way: only the mean influences the motion of the solitary wave, not
vice-versa.  This problem was first considered theoretically and
experimentally in the unidirectional case where a solitary wave and
either a rarefaction wave (RW) mean flow or a dispersive shock wave
(DSW) mean flow exhibited an overtaking interaction
\cite{maiden_solitonic_2018}.  Depending on the conditions, the
solitary wave was observed to be either transmitted through or trapped
by the mean flow as $t \to \infty$.  The key mathematical insight was
the determination of the solitary wave limit of the Whitham modulation
equations---a quasi-linear system of first order equations that
describe the slow evolution of nonlinear wavetrains
\cite{whitham_linear_1999}---and its subsequent diagonalization in
terms of Riemann invariants.  A general approach for obtaining the
solitary wave limit in the case of Hamiltonian partial differential
equations with a local Hamiltonian
can be found in \cite{benzoni-gavage_modulated_2021}.
While Riemann invariants can in principle be obtained for any system
of two quasi-linear, first order PDEs, they are most easily obtained
for completely integrable equations such as the Korteweg-de Vries
(KdV) equation, either directly \cite{whitham_non-linear_1965} or
using the finite gap method
\cite{flaschka_multiphase_1980,a_m_kamchatnov_nonlinear_2000}.  But
Whitham theory can be developed for integrable and non-integrable
equations alike \cite{el_dispersive_2016-1}.  A recent paper has
obtained the diagonalization, i.e., the Riemann invariants, for the
non-integrable Benjamin-Bona-Mahony (BBM) equation
\cite{benjamin_model_1972} by deriving an exact representation of the
solitary wave limit of the BBM-Whitham modulation equations
\cite{gavrilyuk_singular_2021}.

One of our main results in this paper is the derivation of the
modulation solitary wave limit for the apparently non-integrable
Serre-Green-Naghdi (SGN) equations
\cite{serre_contribution_1953,su_korteweg-vries_1969,green_theory_1974,green_derivation_1976}
describing long, strongly nonlinear free surface waves on a flat
bottom in a fluid of constant density.  The Hamiltonian structure of
this system is non-local
\cite{gavrilyuk_generalized_2001,li_hamiltonian_2002}, so the generic
results of \cite{benzoni-gavage_modulated_2021} cannot be used.  The SGN system shares
many properties with the BBM equation: they have similar dispersion
relations and similar nonlocal Hamiltonian structure
\cite{olver_hamiltonian_1980}. Compared to the BBM equation, which is
a unidirectional, scalar equation, the SGN system is a Galilean
invariant system containing more physics than the BBM equation. The
Whitham modulation equations for the SGN equations were obtained in
\cite{el_unsteady_2006}. Their hyperbolicity was proven in
\cite{tkachenko_hyperbolicity_2020}.  Being second order in time, the
SGN equations support bi-directional wave propagation and the
resulting solitary wave limit quasi-linear system of the SGN-Whitham
equations is third order.  In this work, we prove the existence of Riemann
invariants if and only if simple waves in the decoupled mean flow
equations are considered.

The analysis of unidirectional wave-mean flow interaction has been
carried out for solitons and RWs, DSWs for a general class of
unidirectional nonlinear dispersive wave equations in
\cite{maiden_solitonic_2018} by analyzing the solitary wave limit of
the Whitham modulation equations.  In the same paper, the approach was
applied to the conduit equation and compared with experiments on
viscous fluid conduits.  The approach was further refined and applied
to the KdV equation \cite{ablowitz_solitonmean_2023} and the
modified-KdV equation \cite{sande_dynamic_2021} where, in addition,
the kink (monotone, heteroclinic travelling wave solution) serves as
either the wave or the mean flow.  The extension to oblique,
two-dimensional line solitons interacting with RWs and DSWs was
developed for the Kadomtsev-Petviashvili equation
in \cite{ryskamp_oblique_2021}.  A similar analysis of soliton-mean field
interaction was extended to the bi-directional case of the defocusing
nonlinear Schr\"odinger (NLS) equation where, in addition to
overtaking interactions, RWs and DSWs experience head-on interactions
with solitons \cite{sprenger_hydrodynamic_2018}.

The soliton-mean flow interaction problem for integrable equations has
also been studied using the inverse scattering transform (IST) for the
the KdV equation
\cite{ablowitz_solitons_2018,ablowitz_solitonmean_2023} and the
focusing NLS equation \cite{biondini_soliton_2018} where some solitons
were shown to leave a trailing ``wake'' inside the DSW after passing
through.  Other analytical approaches include perturbation theory
\cite{ablowitz_solitonmean_2023}, the Darboux transformation
\cite{mucalica_solitons_2022-1}, both for KdV soliton-RW interaction,
and a Hamiltonian formulation of the problem that borrows ideas from soliton
perturbation theory and Whitham theory to obtain an approximate,
analytical description of solitary wave-mean flow interaction for a
generalized KdV equation \cite{kamchatnov_propagation_2023} and the
NLS equation \cite{ivanov_motion_2022,kamchatnov_hamiltonian_2024}.

In this paper, we utilise the derived SGN-Whitham equations in the
solitary wave limit to analytically describe the  head-on
and overtaking interaction of a solitary wave and a RW mean flow.  The
results are then extended to solitary wave-DSW mean flow interaction.
One important finding in this paper is that the original approach
proposed in \cite{maiden_solitonic_2018} and utilised elsewhere
\cite{kamchatnov_propagation_2023} that rely upon the DSW fitting
method and its conjugate wavenumber/dispersion relation
\cite{el_resolution_2005} to obtain the soliton Riemann invariant is
only approximate.  By obtaining the exact representation of the
solitary wave limit of the SGN-Whitham equations and corresponding
Riemann invariants for simple wave mean flows, we are able to prove
that the conjugate wavenumber/dispersion yields a second order in
amplitude accurate prediction for solitary wave motion through a RW.  It also provides a second order accurate prediction for the DSW's solitary wave edge.
Careful numerical simulations of the SGN equations agree with these
findings.  While this is an improvement to the first order accurate,
weakly nonlinear KdV approximation, it identifies a limitation of the
DSW fitting approach.

The plan of the paper is as follows. In section \ref{SGN
  presentation}, we give a detailed presentation of the SGN equations and their periodic traveling wave solutions are presented in section \ref{periodic TWs}.
In section \ref{Modulation equations}, the modulation equations are
given both in the mass Lagrangian coordinates and in Eulerian
coordinates. The reason is that the solitary wave limit is
mathematically easier to carry out in Lagrangian coordinates, while
the physical interpretation is easier in Eulerian coordinates. In
section \ref{Solitary wave tunnelling}, we study the interaction of
solitary waves with RWs and DSWs. In particular, the transmission and
trapping effect is studied. The closed-form analytical results are in
good agreement with the numerical ones for the SGN equations. Finally,
the main technical details (the different forms of the modulation
equations, the passage to the solitary limit, numerical method, etc.) are given in four
Appendices.

\section{Serre--Green--Naghdi equations}
\label{SGN presentation}

The SGN equations over a flat bottom approximating the free-surface
Euler equations in the long wave limit are
\cite{serre_contribution_1953,su_korteweg-vries_1969,green_theory_1974,green_derivation_1976}
\begin{align}
    &h_t+(h u)_x = 0,   \label{eq:186a} \\
    &u_t+uu_x+ g h_x=\frac{1}{h}\left(\frac{h^3}{3}(u_{xt}+uu_{xx}-
      u_x^2)\right)_x , \label{eq:186b}
\end{align}
where $h$ is the total depth, $u$ is the depth-averaged horizontal
velocity, and $g$ is the acceleration due to gravity. In what follows,
we scale independent and dependent variables so that $g=1$.

The SGN equations \eqref{eq:186a}, \eqref{eq:186b} are non-integrable
and represent a fully nonlinear generalization of the classical
Boussinesq equations \cite{lannes_water_2013}. The first equation
\eqref{eq:186a} is the exact equation for conservation of mass, and
the second equation \eqref{eq:186b} can be manipulated into the
equation for conservation of horizontal momentum
\begin{equation}
  \label{eq:196}
  (h u)_t + \left ( h u^2 + \frac{1}{2} h^2 \right )_x =   \left(\frac{h^3}{3} \left ( u_{tx} + u u_{xx} - u_x^2 \right ) \right)_x  .
\end{equation}
An equivalent form of the momentum equation is
\begin{equation}
  (h u)_t + \left( h u^2 + \frac{1}{2} h^2
    +\frac{1}{3}h^{2}\frac{d^2h}{dt^2} \right)_x =  0 , 
\end{equation}
where $\displaystyle\frac{d}{dt}$ is the  material derivative notation, $\displaystyle \frac{d}{dt}=\frac{\partial }{\partial t}+u\frac{\partial }{\partial x}$, and $\displaystyle \frac{d^2}{dt^2}=\frac{d}{dt} \left( \frac{d}{dt} \right)$ is the second material derivative. The conservation of energy manifests in the
additional conservation law
\begin{equation}
  \label{eq:197}
  \left(\frac{1}{2} h \left ( h + u^2 + \frac{1}{3} h^2
      u_{x}^2 \right ) \right)_t + \left( h u\left ( h +
      \frac{1}{2} u^2 + \frac{1}{2} h^2 u_x^2 - \frac{1}{3}
      h^2(u_{xt} + u u_{xx} ) \right )  \right)_x =0.
\end{equation}
Finally, a fourth conservation law (the so-called Bernoulli
conservation law) can be derived. It is usually written for the
variable
$\displaystyle {\cal{K}}=u-\frac{1}{3h}\left(h^3 u_x\right)_x$
representing the tangent component of the fluid velocity at the free
surface \cite{gavrilyuk_kinematic_2015}
\begin{equation}
  {\cal{K}}_{t}+\left( {\cal{K}} u+h
    -\frac{u^{2}}{2}-\frac{1}{2}h_{x}^2 u^2\right)_{x}=0 . 
  \label{Bernoulli_Eulerian}
\end{equation}
A mathematical justification of the SGN model \eqref{eq:186a},
\eqref{eq:186b} can be found in
\cite{lannes_water_2013,makarenko_second_1986}. Recent years have seen
increased activity in both the study of qualitative properties of the
solutions to the SGN equations and in the development of numerical
discretization techniques
\cite{le_metayer_numerical_2010,li_high_2014,favrie_rapid_2017,gavrilyuk_numerical_2024,gavrilyuk_2d_2024}.

The Lagrangian for \eqref{eq:186a}, \eqref{eq:186b}, where the mass
conservation law \eqref{eq:186a} is considered a constraint, is given
as \cite{gavrilyuk_generalized_2001}
\begin{equation}
  {\cal L}=\int_{-\infty}^{+\infty}
  h\left(\frac{u^{2}}{2}+\frac{1}{6}\left(\frac{dh}{dt}\right)^2
    -\frac{h}{2}\right) dx . 
  \label{Lagrangian_SGN_Euler}
\end{equation}
In order to obtain analytically tractable expressions, we will make
use of mass Lagrangian coordinates introduced as follows. Let $h_0(x)$
be the initial position of the free surface, and $Y$ be the classical
Lagrangian coordinate. The mass Lagrangian coordinate $q$ is defined
as $\displaystyle q=\int_0^Y h_0(s)ds$. Let $x=x(t,q)$ be the
trajectories of fluid particles.  Since the mass conservation law is
in the form $\displaystyle h\frac{\partial x}{\partial q}=1$, the
Lagrangian \eqref{Lagrangian_SGN_Euler} can be transformed to
\begin{equation}
  \tilde {\cal L}=\int_{-\infty}^{+\infty}\left(\frac{u^2}{2}-\tilde
    e(\tau, \tau_t)\right)dq , 
  \label{basic_lagrangian} 
\end{equation}
where $\tau=1/h$ and the potential $\tilde e (\tau, \tau_t)$ is
\begin{equation}
  \tilde e (\tau, \tau_t)=\frac{1}{2\tau}-\frac{1}{6}\left(\frac{\partial}{
      \partial  t} \left(\frac{1}{\tau}\right)\right)^2 =
  \frac{1}{2\tau} -\frac{\tau_t^2}{6\tau^4}. 
  \label{SGN_energy_mass_coordinates}
\end{equation}
For a general potential $\tilde e(\tau, \tau_t)$, the Euler-Lagrange
equations for \eqref{basic_lagrangian} are \cite{gavrilyuk_large_1994}:
\begin{equation}
  \tau_{t}- u_{q}=0,\quad u_{t}+p_{q}=0, 
\label{basic_dispersion_0}
\end{equation}
where the pressure $p$ is defined by
\begin{equation}
  p=-\frac{\delta \tilde e}{\delta \tau} = -\left(\frac{\partial
      \tilde e}{\partial \tau}-\frac{\partial }{\partial t}
    \left(\frac{\partial \tilde e}{\partial \tau_t}\right)\right). 
  \label{definition_pressure_0}
\end{equation}
System \eqref{basic_dispersion_0} is reminiscent of the $p$-system for
the barotropic Euler equations. However, in our case, the pressure $p$
defined by \eqref{definition_pressure_0} depends not only on $\tau$,
but also on its first and second temporal derivatives. The variable
$\tau$ is the analogue of the specific volume for the corresponding
Euler equations. As a consequence of Noether's theorem, the
variational formulation implies the conservation of energy and
Bernoulli equation
\begin{align}
  \label{energy_Lagrangian_0}
  \left( \frac{u^2}{2}+\varepsilon\right)_t+\left( pu\right)_q = 0, \quad
  \varepsilon = \tilde e- &\tau_t \tilde e_{\tau_t}, \\
  \label{Bernoulli_Lagrangian_0}  
  \left(\tau u-\tau_q\frac{\partial \tilde
  e}{\partial\tau_t}\right)_t-\left(\frac{u^2}{2}-\tau p-\tilde
  e\right)_q &= 0. 
\end{align}
that are the analogues of eqs.~\eqref{eq:197} and
\eqref{Bernoulli_Eulerian}.  The conservation laws
\eqref{basic_dispersion_0}, \eqref{energy_Lagrangian_0},
\eqref{Bernoulli_Lagrangian_0} are averaged to obtain the modulation
equations in Lagrangian coordinates in Appendix \ref{Appendix_A}.

\section{Periodic travelling wave solutions}
\label{periodic TWs}

We now present the form of periodic travelling wave solutions in both
Eulerian and Lagrangian coordinates.  Several parameterisations of
the four dimensional family of periodic travelling waves will be
presented.

\subsection{Eulerian coordinates}

The travelling wave solutions $h(x,t) = h(\xi) = h(x-ct)$,
$u(x,t) = u(\xi) = u(x-ct)$ to the SGN equations \eqref{eq:186a},
\eqref{eq:186b} are specified by
\begin{align}
  \label{ODE_serre}
  &\left(\frac{dh}{d\xi}\right)^2 = \frac{3}{h_1 h_2
    h_3}(h-h_1)(h-h_2)(h_3 - h)=\frac{3}{h_1 h_2
    h_3}P(h), \\
  \label{ODE_serre_velocity}
  &u=c {\color{blue}-} \sigma \frac{\sqrt{h_1h_2h_3}}{h},\quad
    \sigma = \pm 1,
\end{align}
where, similar to the defocusing nonlinear Schr\"odinger equation, the
wave is characterised by four independent parameters $h_1$, $h_2$,
$h_3$ (the roots of the third order polynomial $P(h)$), the travelling
wave velocity $c$, and $\sigma=+1$ ($\sigma=-1$) corresponds to the
fast (slow) waves.  Here, the depth variations in the wave occur in
the interval $\left[h_2, h_3\right]$, for $ h_1 < h_2 < h < h_3$ with
the amplitude $a=h_3-h_2$. It is assumed that there are no vacuum
points for non-trivial periodic travelling wave solutions: $h_1>0$.
The ODE in \eqref{ODE_serre} can be integrated to obtain
\begin{equation}
  \label{serre_TW}
  h = h_2 + (h_3- h_2)\mathrm{cn}^2\left
    (\frac12\sqrt{\frac{3(h_3 - h_1)}{h_1 h_2 h_3}} \xi
    , m\right), \quad m=\frac{h_3 - h_2}{h_3 - h_1},
\end{equation}
where $\mathrm{cn}$ is the Jacobi cosine elliptic function
\cite{olver_nist_2024}.  We express the physical parameters (amplitude
$a$, wavenumber $k$, and the period averages of depth $\hb$ and
velocity $\ub$) in terms of the basic parameter set
$(h_1, h_2, h_3, c)$ as 
\begin{equation}
  \label{serre_mean}
  \begin{split}
    &a = h_3 - h_2, \quad k = \sqrt {\frac{3(h_3 -
        h_1)}{h_1 h_2 h_3}} \frac{\pi}{2 \K(m)}, \quad
    \overline h = h_1 +
    (h_3 - h_1)\frac{\E(m)}{\K(m)} , \\
    & \overline{h (u-c)} =
    {- \sigma} \sqrt{h_1 h_2 h_3}, \quad
    \ub = c  {- \sigma} \sqrt{h_1 h_2 h_3}\; \frac{\Pi \left (1-\frac{h_2}{h_3} ,m \right)}{h_3 \K(m)},
 \end{split}
\end{equation}
where $\K$, $\E$ and $\Pi$ are the complete elliptic integrals of the
first, second, and third kinds, respectively \cite{olver_nist_2024}.  We introduce the phase
\begin{equation}
  \label{eq:phase}
  \theta = k \xi = kx - \omega t,
\end{equation}
so that the traveling wave is $2\pi$-periodic in $\theta$: $h(\xi) = h(\theta/k) = h((\theta+2\pi)/k)$, $u(\xi) = u(\theta/k) = u((\theta + 2\pi)/k)$. 
The wavelength of the travelling wave $L$ is related to the wavenumber
$k$ by $L=2\pi/k$.

\begin{figure}[h]
	\centering
	\includegraphics[width=6cm]{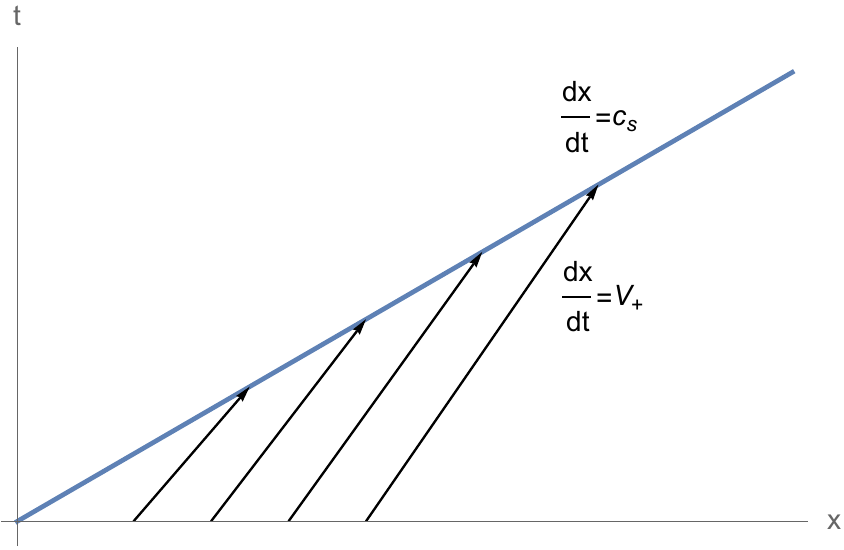}
    \hspace{1cm}
    \includegraphics[width=6cm]{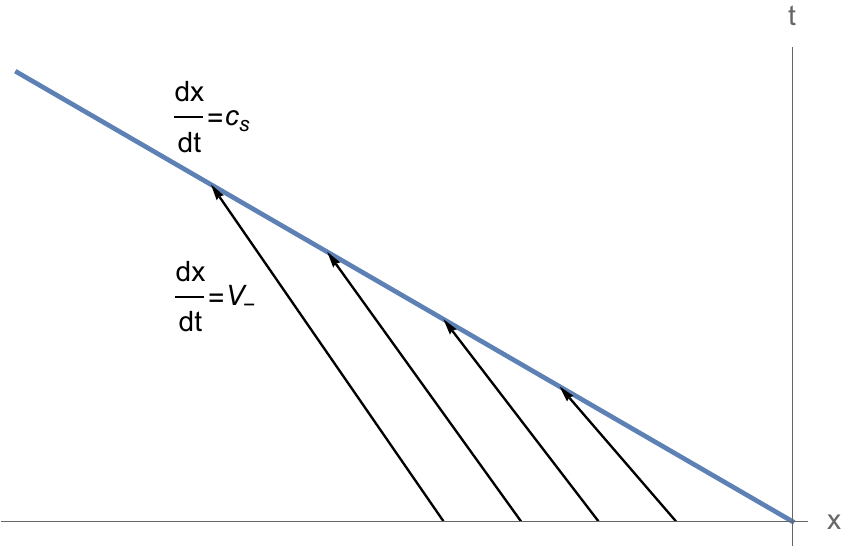}
    \caption{Relations \eqref{serre_speed-amp} between the velocity of
      a fast (slow) solitary wave and the corresponding long-wave
      velocity $V_+$ ($V_-$) are represented schematically on the left
      (right).}
	\label{fig:fast and slow}
\end{figure}

The solitary wave limit of \eqref{serre_TW} is achieved when
$h_2 \to h_1$ so that $m \to 1$. Its explicit form is
\begin{equation}
  \label{eq:198}
  h(x,t)= \hb + a \,\mathrm{sech}^2 \left ( \frac{\sqrt{3a}}{\hb
      \sqrt{\hb + a}} (x - c t)\right ),\quad u(x,t)= \ub + \sigma
  \sqrt{\hb + a} \left ( 1 - \frac{\hb}{h(x,t)} \right ) . 
\end{equation}
Fast ($\sigma=+1$) and slow ($\sigma=-1$) elevation solitary waves
propagate on the background $h=\hb, u=\ub$, and are characterised by
the speed-amplitude relation
\begin{equation}
	\label{serre_speed-amp}
	c=c_s(a, \bar  h, \ub) \equiv \ub +\sigma \sqrt{\hb +a}.
\end{equation} 
Fast solitary waves move faster than the dispersionless long-wave
velocity $V_+ = \ub + \sqrt{\hb}$, and slow solitary waves move slower than  the dispersionless long-wave
velocity $V_- = \ub - \sqrt{\hb}$ (see Figure \ref{fig:fast and slow}).

In the opposite, harmonic, limit, $h_2 \to h_3$ so that $m \to 0$,
yields a small-amplitude linear wave characterised by the dispersion
relation $\omega_0(k)$ (frequency-wavenumber relation) for linear
waves propagating on the background $ ({\hb},\ub)$
\begin{equation}
	\label{serre_disp_rel}
	\omega = k c= \omega_0(k,\hb, \ub)\equiv k \ub + \sigma  
	k \sqrt{\frac{ \hb}{1+ \hb^2k^2/3}} .
\end{equation} 

%

\subsection{Lagrangian  coordinates}

Consider now SGN's travelling wave solutions in the co-moving mass Lagrangian
coordinate $\zeta = q - \tilde{c}t$ instead of the Eulerian coordinate $\xi = x - ct$:
$h=h(\zeta)$, $u=u(\zeta)$. Since
$\displaystyle\frac{d\zeta}{d\xi}=h$, the equation for traveling
waves is
\begin{equation}
  \label{ODE_lagrangian}
  \left(\frac{dh}{d\zeta}\right)^2 = \frac{3}{h_1 h_2
    h_3}\frac{P(h)}{h^2}.
\end{equation}
The velocity $u$ is found from the relation
\begin{equation}
  \tilde c \,\tau+\; u={\rm cst}, \quad {\rm with}\quad
  \tau=\frac{1}{h}.
\end{equation}
The wave velocity $\tilde c$ is related to the roots of the polynomial
$P(h)$ by the formula (see Appendix \ref{Appendix_A})
\begin{equation} 
  \tilde{c}^2=h_1h_2h_3 .
\end{equation}

\section{Modulation equations and their solitonic reduction}
\label{Modulation equations}

The SGN equations possess all the necessary prerequisites for the
application of Whitham averaging.  They support a family of
$2\pi$--periodic travelling wave solutions $h(\theta/k), u(\theta/k)$ specified by
\eqref{serre_TW}, \eqref{eq:phase} and characterized by the four independent parameters
$h_1, h_2, h_3, c$ while admitting four independent conservation laws
\eqref{eq:186a}, \eqref{eq:186b}, \eqref{eq:197} and
\eqref{Bernoulli_Eulerian}.

The modulation equations for the SGN equations in Eulerian coordinates
are given in \cite{el_unsteady_2006,tkachenko_hyperbolicity_2020}.
According to the general averaging procedure
\cite{whitham_non-linear_1965}, the modulation system for the SGN
equations can be obtained by period averaging any three conservation
laws, such as mass \eqref{eq:186a}, momentum \eqref{eq:196} and energy
\eqref{eq:197}, over the periodic family \eqref{serre_TW}, and
augmenting them by the wave conservation equation $k_t +
(kc)_x=0$. Doing so results in
\begin{equation}
  \label{serre_gen_mod}
  \begin{split}
    &\overline{h}_t+(\overline{h u})_x = 0, \\
    &( \overline{h u} )_t+\left( \ \tfrac 12 \overline{h^2}+\overline{h 
      u^2}- \tfrac{1}{3}  \overline{h^3\left( (u-c) 
      u_{\xi \xi} 
      -u_{\xi\xi}^2\right)} \ \right)_x=0, \\
    &\left(\frac{1}{2} \overline{h \left ( h + u^2 + \frac{1}{3} h^2
      u_{\xi}^2 \right )} \right)_t + \left( \overline{h u\left( h +
      \frac{1}{2} u^2 + \frac{1}{2} h^2 u_\xi^2 - \frac{1}{3}
      h^2(u-c) u_{\xi\xi} ) \right)}  \right)_x =0 ,\\
    &k_t + \left ( k c \right )_x=0 .
  \end{split}
\end{equation}
The averages in \eqref{serre_gen_mod} are evaluated using the general
definition
\begin{equation}
  \overline {f(h)} =  \dfrac{\displaystyle\int_{h_2}^{h_3}
    \dfrac{f(h)dh}{\sqrt{P(h)}}}{\displaystyle\int_{h_2}^{h_3}\dfrac{dh}{\sqrt{P(h)}}
  },
  \label{averaging_f}
\end{equation} 
where $f(h)$ is any function of $h$.


The system \eqref{serre_gen_mod} is consistent with the averaged Bernoulli conservation law~\eqref{Bernoulli_Eulerian}:
\begin{equation}
\left( \ \overline{u}+\tfrac{1}{6} \overline{h^2u_{\xi
        \xi}} \ \right)_t + 
  \left( \ \tfrac12 \overline{u^2}+\overline{h}-\tfrac12
       \overline{h^2 
      \left(\left(\tfrac23 u- c\right) u_{\xi
          \xi}-u_\xi^2\right)}  \
  \right)_x=0 ,
\end{equation}
which can be used instead of any of the averaged conservation laws
\eqref{serre_gen_mod}, e.g., instead of the wave conservation equation,
yielding an equivalent modulation system.  This equivalence is proved
in the mass Lagrangian coordinates in Appendix~\ref{Appendix_A}.

As a result, the system \eqref{serre_gen_mod} can be explicitly
represented in canonical quasilinear form for the state vector
${\bf b}= (h_1(x,t), h_2(x,t), h_3(x,t), c(x,t))^T$; see
Appendix~\ref{Appendix B}.  However, due to the lack of integrable
structure for the SGN equations, the associated modulation system
cannot be reduced to a diagonal form.  This makes the analysis of its
properties---hyperbolicity, genuine nonlinearity, simple waves,
etc.---difficult.  The weakly nonlinear regime of the modulation
equations was studied in \cite{el_unsteady_2006}. The hyperbolicity of
the modulation equations was proven in
\cite{tkachenko_hyperbolicity_2020}.  However, their full analytical
study remains a difficult task. This is why we study here a reduction
of the full modulation system in the limit of waves of large
wavelength (the solitary wave limit), allowing us to present some
analytical results.  The difficulty of obtaining the solitary wave
limit of the modulation equations is that the phase equation $k_t+(ck)_x=0$
is degenerate in this singular limit. One possibility is to pass to
the limit in the wave action conservation law, an exact conservation
law of the full SGN-Whitham system. Such a method was, in particular,
exploited in \cite{gurevich_nonlinear_1990} for the KdV equation, in
\cite{gavrilyuk_singular_2021} for the BBM equation and in
\cite{benzoni-gavage_modulated_2021} for Hamiltonian systems of
Euler-Korteweg or dispersive Eulerian type (so-called second
gradient fluids).  The wave action integral for the SGN-Whitham
equations was already obtained in \cite{gavrilyuk_large_1994} in a
general setting that includes the SGN-Whitham equations by using mass
Lagrangian coordinates.  We use this result to first find the solitary
wave limit in mass Lagrangian coordinates and then re-express the
limit equations in Eulerian coordinates (see Appendix
\ref{Appendix_A}). In the mass Lagrangian coordinates, one has only to
replace the polynomial $P(h)$ by $P(h)/h^2$ when computing averages (see
\eqref{ODE_lagrangian}):
\begin{equation}
  \overline{f(h)} = \dfrac{\displaystyle \int_{h_2}^{h_3}
    \dfrac{hf(h)dh} {\sqrt{P(h)}}}
  {\displaystyle\int_{h_2}^{h_3}\dfrac{hdh} {\sqrt{P(h)}}}. 
  \label{averaging_Lagrangian_f}
\end{equation} 
To double-check our computation of the solitary wave limit in mass
Lagrangian coordinates, we also compute the limit directly in Eulerian
coordinates (see \ref{Appendix B}) using a symbolic package.

A straightforward analysis shows that in the solitary wave limit, one
has $\hb=h_1$, $\overline{h^2} = h_1^2 = \hb^2$, $\overline{1/h}=1/h_1=1/\hb$, and the average of the
mass and momentum equations in mass Lagrangian coordinates
\eqref{basic_dispersion_0} are thus the classical Saint--Venant
equations describing hydrostatic (dispersionless) shallow water
equations
\begin{equation}
	\label{eq:sol_limit_mass_momentum_lagrangian}	
    \left(\frac{1}{\hb} \right)_t-\ub_q=0, \quad
    \ub_t-\left(\frac{\hb^2}{2}\right)_q=0. 
\end{equation}
This decoupling of the dispersionless limit equations for the
evolution of the mean flow in the solitary wave limit is a general
property of dispersive hydrodynamics
\cite{el_resolution_2005,hoefer_shock_2014,el_dispersive_2016-1}. It
is also physically intuitive: the effect of a single soliton on a
large-scale mean flow is negligibly small.

A non-trivial amplitude equation comes from the solitary wave limit of
the wave action conservation equation.  For general periodic travelling
waves in mass Lagrangian coordinates, the wave action equation is
(see Appendix \ref{Appendix_A})
\begin{equation}
  \label{action_1} 
  \left(\tilde c  \left(\frac{\overline{h_\zeta^2}}{3k}
      +\frac{\overline{\tau^2} -\overline{\tau}^2}{k}\right)\right)_t
  +\left(\tilde c^2\,  \frac{\overline{\tau^2}
      -\overline{\tau}^2}{k}\right)_q=0.
\end{equation} 
The solitary wave limit ($k\rightarrow 0$) is singular, because
$\overline{h_\zeta^2}\rightarrow 0$ and
$\overline{\tau^2}-\overline{\tau}^2\rightarrow 0$. In Appendix
\ref{Appendix_A}, we show that equation \eqref{action_1} limits to
\begin{equation}
  \label{eq:sol_limit_amplitude_lagrangian}
  F(n,\hb)_t+G(n,\hb)_q=0, 
\end{equation} 
with 
\begin{align}
	\label{eq:FF} 
&F(n,\hb) =\frac{\hb^{\frac{3}{2}}}{\sqrt{1-n}}
                     \left(\frac{(6-2n)\sqrt{n}}{3(1-n)} +{\rm
                     ln}\left({\frac{1-\sqrt{n}}{1+\sqrt{n}}}\right)\right),\\ 
	\label{eq:GG}  
  &G(n,h_1)=\sigma \frac{\hb^{3}}{1-n}
    \left(\frac{\sqrt{n}}{1-n}+\frac{1}{2}{\rm
    ln}\left({\frac{1-\sqrt{n}}{1+\sqrt{n}}}\right)\right) . 
\end{align}
The quantity $n$ is related to the solitary wave amplitude $a=h_3-h_1$
as
\begin{equation}
  \label{eq:n_soli}
  n=\frac{a}{\hb+a} .
\end{equation}
The characteristic velocity corresponding to equation
\eqref{eq:sol_limit_amplitude_lagrangian} is
\begin{equation}
  \label{eq:lambda}
  \lambda=\frac{G_n}{F_n}=\sigma\frac{\hb^{3/2}}{\sqrt{1-n}}=\sigma
  \hb\sqrt{\hb+a} = \tilde{c}_s.
\end{equation}
The velocity $\tilde{c}_s$ is related to the Eulerian velocity of solitary waves $c_s$  by \eqref{serre_speed-amp}.

In Eulerian coordinates, the system
\eqref{eq:sol_limit_mass_momentum_lagrangian},
\eqref{eq:sol_limit_amplitude_lagrangian} becomes
\begin{align}
\label{eq:hb}
  &\hb_t+(\hb \ub)_x=0, \\
  \label{eq:ub}
  &\ub_t+\ub\,\ub_x+ \hb_x=0,\\
  &z_t+ \left(\ub + \sigma \sqrt{\hb(1+z^2)} \right) z_x + \sigma \frac{3  \left(z^2+1\right)^{3/2}}{2z} \, \frac{
    z \sqrt{z^2+1} -\sinh ^{-1}(z)}{2
    z \sqrt{z^2+1}- \sinh
    ^{-1}(z)}\,  \frac{\hb_x}{\sqrt{\hb}}
\label{eq:z}\\
    &\hspace{4cm}-\frac{\sqrt{z^2+1}}{2z}\, \frac{3z+2z^3-3\sqrt{z^2+1}\sinh
    ^{-1}(z)}{2z\sqrt{z^2+1}-\sinh
    ^{-1}(z)}\,  \ub_x=0. \nonumber
\end{align}
where $z^2=a/\hb$\, and \, $\sinh^{-1}(z)=\ln(z+\sqrt{z^2+1})$. We
call the system \eqref{eq:hb}--\eqref{eq:z} the solitonic modulation
system. We stress that the solitonic modulation system is an {\it
  exact reduction} of the full SGN-Whitham modulation system.

The first two equations of the solitonic modulation system are the
Saint-Venant (shallow water) equations written in Eulerian
coordinates. They are completely decoupled from the third equation for
$z$.  Other than classical fast and slow surface waves of the shallow
water equations with characteristic velocities
$V_{\pm}=\ub\pm \sqrt{\hb}$, the characteristic velocity corresponding
to the amplitude equation for the dimensionless amplitude $z^2$ also
represents fast and slow solitary waves propagating with the third
characteristic velocity $c_s=\ub + \sigma \sqrt{\hb(1+z^2)}$, the
solitary wave velocity \eqref{serre_speed-amp}.  By direct
verification, we find that the system is strictly hyperbolic
$V_-<V_+< c$, if $\sigma=+1$, $\hb > 0$, and $z > 0$ or $c<V_-<V_+$,
if $\sigma=-1$, $\hb > 0$, and $z > 0$ and the corresponding
eigenfields are genuinely nonlinear in the sense of Lax.

The solitonic modulation system \eqref{eq:hb}--\eqref{eq:z} inherits
two Riemann invariants, those of the decoupled shallow water equations
\eqref{eq:hb}, \eqref{eq:ub}. However, the third Riemann invariant
does not exist, i.e. the system cannot be fully diagonalized (see
Appendix~\ref{Appendix C} for the proof). Nevertheless, given the
solution of \eqref{eq:hb} and \eqref{eq:ub} (obtained for example by
the hodograph transform or a simple wave reduction), eq.~\eqref{eq:z}
can then be integrated using the method of characteristics.  Moreover,
if one of the shallow water Riemann invariants is constant (the simple
wave reduction), there are just two equations left, e.g.~\eqref{eq:ub}
and \eqref{eq:z}, and so the amplitude equation \eqref{eq:z} can also
be diagonalized in that case. The resulting extra Riemann invariant
$Q(x,t)$ plays the role of an adiabatic invariant for solitary
wave-mean flow interaction and determines the transmission (tunnelling)
and trapping conditions for the interaction of a solitary wave with a
RW or DSW mean flow generated by step initial data (Riemann) problems.

\section{Solitary wave transmission and trapping}
\label{Solitary wave tunnelling}

The expression ``solitary wave tunneling'' comes from quantum mechanics
and conveys the possibility of a quantum particle passing through a classically impenetrable potential 
barrier. In the context of the SGN equations, the solitary wave plays
the role of the quantum particle.  More precisely, for a given solitary wave
amplitude, we are able to use the solitary wave limit of the
SGN-Whitham equations to analytically determine if the solitary wave
passes (transmits or tunnels) through a RW or its counterpart, a DSW.
Both a RW and a DSW represent a continuous transition between two constant mean
flows and therefore solitary wave-mean flow transmission occurs when a
solitary wave, initiated on one side of a RW or DSW, emerges on the
other side.  Otherwise, we say that the solitary wave has been
trapped.



The system~\eqref{eq:hb}, \eqref{eq:ub} is independent of $z$ and can
be diagonalized. The corresponding Riemann invariants are:
\begin{equation} \label{riem_SV}
  r_+ = \ub + 2\sqrt{\hb},\quad r_- = \ub - 2\sqrt{\hb},
\end{equation}
yielding
\begin{align}
  &(r_+)_t + V_+ (r_+)_x=0,\quad V_+ = \ub + \sqrt{\hb} =
    \frac{3r_++r_-}{4}, \\
  &(r_-)_t + V_- (r_-)_x=0,\quad V_- = \ub - \sqrt{\hb} =
    \frac{r_++3r_-}{4}.
\end{align}
This system has two simple wave solutions: 
\begin{equation}\label{eq:simple}
  r_{-\mu}=\ub-2\mu \sqrt{\hb} = {\rm cst},
  \quad V_{\mu} =  x/t,
\end{equation}
where $\mu=+1$ for the fast wave and $\mu=-1$ for the slow wave.

\subsection{Exact simple wave Riemann invariants}
\label{sec:exact-simple-wave}

To study the interaction of the solitary wave with the simple wave
\eqref{eq:simple}, we hold one of the Riemann invariants $r_{-\mu}$
constant, yielding a relation between $\hb$ and
$\ub$~\eqref{eq:simple} \cite{sprenger_hydrodynamic_2018}.  The
reduced solitonic modulation system is thus obtained by
substituting~\eqref{eq:simple} in~\eqref{eq:hb}, \eqref{eq:z} (or
equivalently \eqref{eq:ub}, \eqref{eq:z}), yielding the system of two
equations:
\begin{align}
\label{eq:hb2}
  &\hb_t+\left(r_{-\mu}+3\mu\sqrt{\hb}\right)\hb_x=0,\\
  \label{eq:z2}
  & z_t+\left(r_{-\mu}+2\mu\sqrt{\hb}+ \sigma \sqrt{\hb(1+z^2)}
    \right) z_x + \sigma g_{\sigma \mu}(z)\frac{\hb_x}{\sqrt{\hb}}
    =0, 
\end{align}
with 
\begin{equation}
  g_{\sigma \mu}(z)=\frac{3  \left(z^2+1\right)^{3/2}}{2z} \, \frac{
    z \sqrt{z^2+1} -\sinh ^{-1}(z)}{2
    z \sqrt{z^2+1}- \sinh^{-1}(z)}-\sigma \mu \frac{1+z^2}{2z} \,
  \frac{(3z+2z^3)(1+z^2)^{-1/2}-3\sinh 
    ^{-1}(z)}{2z\sqrt{1+z^2}-\sinh
    ^{-1}(z)}
\end{equation}
Since now this is a quasi-linear system of just two equations, it is
diagonalizable with one Riemann invariant manifestly $\hb$.  By making
the simple-wave ansatz $z=z(\hb)$ in eqs.~\eqref{eq:hb2} and
\eqref{eq:z2}, the second Riemann invariant can be obtained as the
constant of integration of the ODE
\begin{equation}
  \label{ODE_q}
  \frac{dz}{d\hb} + \frac{g_{\sigma \mu}(z)}{\hb
    \left(\sqrt{1+z^2}-\sigma \mu\right)} =0. 
\end{equation}
The integral of \eqref{ODE_q} can be written in the form
\begin{equation}
  \label{eq:f}
  Q_{\sigma \mu}(\hb,z) = \hb \exp(f_{\sigma \mu}(z)),\quad 
  f_{\sigma \mu}(z) = \int^z \frac{\sqrt{1+s^2}-\sigma\mu}{g_{\sigma \mu}(s)}\, ds.
\end{equation}
One can see that the Riemann invariant $Q_{\sigma \mu}$ \eqref{eq:f} only depends on
the sign $\sigma \mu$, which determines the interaction type.  The
value $\sigma \mu=+1$ corresponds to the overtaking interaction
between a fast solitary wave and a fast simple wave (or equivalently
slow-slow waves) whereas $\sigma \mu=-1$ corresponds to the head-on
interaction between a fast solitary wave and a slow simple wave (or
equivalently slow-fast waves).

Note that in the small amplitude limit $z \to 0$, we obtain the
following expansion for the Riemann invariant in the case of an
overtaking interaction ($\sigma \mu=+1$)
\begin{equation}
  \label{eq:qp}
  Q_+(\hb,z) = \hb \left(1+\frac{z^2}{2}-\frac{z^4}{48}
    -\frac{61}{1440} z^6+ O(z^8)\right). 
\end{equation}
As we will now show, this expression is asymptotically equivalent, to
order $\mathcal{O}(z^4)$, to the Riemann invariant obtained by the DSW
fitting method.

\subsection{Approximate simple wave Riemann invariants: the DSW fitting
  method}
\label{sec:appr-simple-riem}

An efficient approach to obtain the solitary wave limit of the Whitham
equations can be deduced from the dispersive shock wave (DSW) fitting
method \cite{el_resolution_2005}. This method enables the
determination, under certain assumptions, of the harmonic and solitary
wave edges of a DSW directly, bypassing the derivation and asymptotic
analysis of the full Whitham modulation system. The DSW fitting method
has been successfully applied to many dispersive hydrodynamic systems,
both integrable and non-integrable, see e.g.,
\cite{el_undular_2005,el_unsteady_2006,el_theory_2007,esler_dispersive_2011,lowman_dispersive_2013-1,hoefer_shock_2014,congy_dispersive_2021,jamshidi_long-wave_2020}. In
cases when the dispersive equation is integrable such as in the KdV,
NLS, and Kaup-Boussinesq equations, the method's results are
consistent with the available exact modulation solutions. For
non-integrable systems, when the exact theory is not available, the
method yields an excellent comparison with direct numerical
simulations of the dispersive equation, beyond the classical weakly
nonlinear KdV approximation. Here, we shall use the part of the DSW
fitting construction pertaining to the determination of the DSW's
solitary wave edge.

A major assumption in the DSW fitting method is that of convexity
(strict hyperbolicity and genuine nonlinearity) of the full Whitham
modulation system. This assumption is not always easily verifiable for
the entire range of parameters involved, so one can say that DSW
fitting generally provides a {\it convex approximation} of the exact
DSW edge dynamics. In particular, it is known that the SGN-Whitham
modulation system exhibits non-convexity for sufficiently large
amplitudes \cite{el_unsteady_2006,hoefer_shock_2014} but nevertheless,
the DSW fitting results agree with direct SGN numerics remarkably well
for a broad range of amplitudes, far beyond the weakly nonlinear KdV
regime.

The DSW fitting method relies on the existence of exact reductions of
the Whitham modulation system in two distinguished limits: the
harmonic limit $a \to 0$ and the solitary wave limit $k\to 0$. The
latter one is our interest here and has the general form of an
equation for the solitary wave amplitude (or a related amplitude type
variable), coupled to the dispersionless limit system for the
large-scale background (the mean flow)
\cite{el_dispersive_2016-1}. In the present context of the SGN-Whitham
equations, the solitary wave reduction is given by the system
\eqref{eq:hb}--\eqref{eq:z}, however, the derivation of the exact form
of the amplitude equation \eqref{eq:z} and further, of the Riemann
invariant $Q_{\sigma \mu}$ in eq.~\eqref{eq:f}---the main technical
hurdle in the analysis of the solitary wave limit of the modulation
system---is not required for the application of the DSW fitting
method.  Instead, the determination of the requisite Riemann invariant
involves only the linear dispersion relation $\omega_0(k, \hb, \ub)$
for waves on the mean background $(\hb,\ub)$ along with the
simple-wave relation $\ub(\hb)$ for the dispersionless limit
equations.  All of this information is readily available with no
additional analysis of the SGN-Whitham modulation system required. In
the context of soliton-mean flow interaction, the Riemann invariant
$Q$ acquires the meaning of an adiabatic invariant
\cite{maiden_solitonic_2018}. This adiabatic invariant also plays a
key role in the solitary wave resolution method
\cite{el_asymptotic_2008,maiden_solitary_2020}, where it is an
analogue of the spectral parameter in the Lax pair associated with
integrable equations in the semi-classical limit.  See also the recent
papers \cite{kamchatnov_propagation_2023,kamchatnov_hamiltonian_2024}
where soliton-mean field interaction has been interpreted in terms of
Hamiltonian mechanics of the motion of a soliton along a large-scale
background.

Note that the DSW fitting construction applies directly to the
overtaking solitary wave-mean field interaction, not the head-on
interaction, due to the unidirectional nature of DSW generation in
Riemann problems.  The DSW fitting method was applied to undular bore
theory for the SGN system in \cite{el_unsteady_2006}, so we shall take
advantage of the results of that work and adapt them to the current
setting of the overtaking solitary wave-mean flow interaction. To be
definite, we consider the fast-fast interaction, $\sigma = \mu =+1$.

For the fast simple-wave mean field we have (cf.~\eqref{eq:hb2})
\begin{equation}
  \label{simple_fast}
  \ub(\hb)=r_-+ 2\sqrt{\hb}, \quad \hb_t+V_+(\hb) \hb_x=0,\quad V_+ =
  r_{-}+3\sqrt{\hb},  
\end{equation}
where $r_-= \ub - 2 \sqrt{\hb} = const$ is the Riemann invariant of
the dispersionless shallow-water  equations.  The fast
branch of the dispersion relation \eqref{serre_disp_rel} for
linearised waves on the simple-wave background \eqref{simple_fast} is
given by
\begin{equation}
	\label{serre_disp_rel-fast}
	 \omega_0(k,\hb, \ub(\hb)) =  k\left(r_{-}+2 \sqrt{\hb} \right) +  
	k \sqrt{\frac{ \hb}{1+ \hb^2k^2/3}} .
\end{equation} 

Solitary wave motion on the simple-wave background is defined by the
conjugate dispersion relation (see \cite{el_resolution_2005} for
details)
\begin{equation}
  \omegat(\kt,\hb)=-i\omega_0 \left(i \kt,\hb, \ub(\hb) \right) =\kt
  \left(r_{-}+2 \sqrt{\hb} \right) +  
  \kt \sqrt{\frac{ \hb}{1- \hb^2 \kt^2/3}},
\end{equation}
where the conjugate wavenumber $\kt$ (essentially the inverse width of the solitary wave) is related to the solitary wave amplitude $a$ by
\begin{equation}
    \frac{\omegat}{\kt} = c_s(a,\hb,\ub(\hb)),
\end{equation}
with $c_s(a,\hb,\ub)=\ub + \sqrt{\hb +a}$ the fast branch of the SGN
solitary wave speed-amplitude relation \eqref{serre_speed-amp}.

Within the fitting approach to solitary wave-mean flow interaction,
the amplitude modulation equation is obtained directly in diagonal
form. The corresponding Riemann invariant $\tilde Q_+(\tilde k, \hb)$
satisfying
\begin{equation}\label{eq:q}
\frac{\partial \tilde Q_+}{\partial t} + C(\tilde Q_+, \hb) \frac{\partial \tilde Q_+}{\partial x} =0, \quad C(\tilde Q_+, \hb) = c_s(a(\tilde k, \hb), \hb, \ub(\hb))
\end{equation}
is found as an integral of the characteristic ODE
\begin{equation}\label{char_ODE}
  \frac{d \kt}{d \hb} = \frac{\omegat_\hb}{V_+(\hb)-\omegat_\kt} =
  \frac{\kt\left( \left(1+\frac{\hb^2 \kt^2}{3}\right)+2
      \left(1-\frac{\hb^2  \kt^2}{3}\right)^{3/2}+1 \right)}{2 \hb
    \left(  \left(1-\frac{\hb^2 \kt^2}{3}\right)^{3/2}-1 \right)}. 
\end{equation}
Equation \eqref{char_ODE} is a \textit{convex approximation}
counterpart to the ODE \eqref{ODE_q} obtained by the exact evaluation
of the solitary wave limit of the SGN-Whitham modulation system.

With the change of variable
\begin{equation}
  \alpha = \frac{1}{\sqrt{1-\frac{\hb^2 \kt^2}{3}}} =
  \sqrt{1+\frac{a}{\hb}} = \sqrt{1+z^2}, 
\end{equation}
eq.~\eqref{char_ODE} reduces to the separable ODE
\begin{equation}
    \frac{d \alpha}{d\hb} = \frac{\alpha \left(\alpha^2-1\right) (4 -\alpha)}{2 \hb \left(1-\alpha^3\right)},
\end{equation}
which is readily integrated with the constant of integration
\begin{equation}
  \label{eq:qfit}
  \tilde Q_+ = \frac{2^{2/5} 3^{21/10}\hb \alpha^{1/2}}{(\alpha
    +1)^{2/5} (4-\alpha)^{21/10}}. 
\end{equation}
The normalization coefficient in \eqref{eq:qfit} is chosen by the
natural requirement that in the zero-amplitude/infinite width limit
$a \to 0$ of the solitary wave \eqref{eq:198}, the Riemann invariant
$\tilde Q_+ \to \hb$ so that equation \eqref{eq:q} degenerates into
the simple wave mean flow equation \eqref{simple_fast} for $\hb$.
Equation \eqref{eq:qfit} is equivalent to eq.~(49) in
\cite{el_unsteady_2006}.

\begin{figure}
  \centering
  \includegraphics[width=6cm]{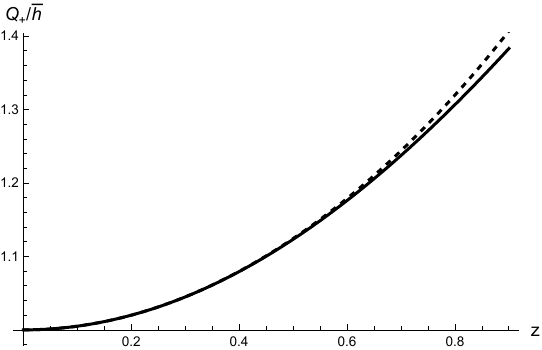} \quad
  \includegraphics[width=6cm]{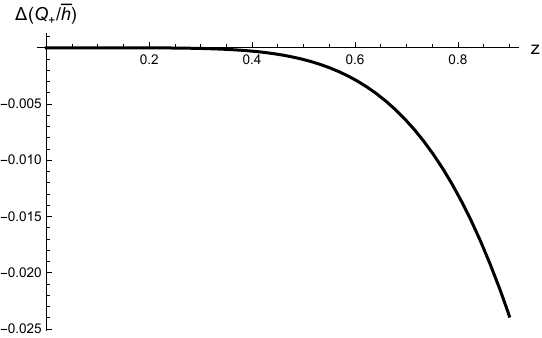} 
  \caption{Left: comparison between the exact Riemann invariant $Q_+$ \eqref{eq:f} (solid line) and  its
    convex approximation $\tilde {Q}_+$\eqref{eq:qfit} (dashed line); Right: the difference $\Delta= (Q_+ - \tilde Q_+)/ \overline h$.    
    }
  \label{fig:fitting_exact_RI}
\end{figure}
One can see that the expression \eqref{eq:qfit} does not coincide with
the rigorous asymptotic result \eqref{eq:f} for $Q_+$ corresponding to
$\mu=\sigma=1$. However, it provides a quite accurate approximation in
the physically relevant range of amplitude-depth ratios
$z^2=a/\hb$. Indeed, the small amplitude expansion of the fitting
approach result \eqref{eq:qfit} is
\begin{equation}
  \label{eq:small_amp}
  \tilde Q_+ = \hb \left ( 1+\frac{z^2}{2}-\frac{z^4}{48}
  -\frac{37}{864} z^6+ {\cal O}(z^8) \right ). 
\end{equation}
The first two terms in this expansion correspond to the Riemann
invariant of the KdV modulation system in the soliton limit
\cite{el_dispersive_2016-1}, i.e., is accurate to $\mathcal{O}(a)$
($\mathcal{O}(z^2)$) as $a \to 0$. But the fitting approach improves
upon the classical KdV prediction since \eqref{eq:small_amp} and the
exact result \eqref{eq:qp} asymptotically agree up to
$\mathcal{O}(a^2)$ ($\mathcal{O}(z^4)$). Even further, the difference
between the two expansions at $\mathcal{O}(a^3)$ ($\mathcal{O}(z^6)$)
is just $\sim 10^{-3} a^3$, i.e. the exact \eqref{eq:f} and convex
approximation \eqref{eq:qfit} curves are in excellent agreement for
physically admissible $a/\hb < a_{\rm max}/\hb \approx 0.8$ \cite{el_unsteady_2006}, see
Fig.~\ref{fig:fitting_exact_RI}. At the same time, the existence of
such a discrepancy poses the important question:  what is the cause of
this $\mathcal{O}(a^3)$ asymptotic discrepancy and what does it mean
for the asymptotic validity of the DSW fitting method, particularly
for non-integrable systems?

\subsection{Riemann problem and transmission condition}

We consider the Riemann problem for the system~\eqref{eq:hb},
\eqref{eq:ub}, \eqref{eq:z}, for which the initial condition is a step
function:
\begin{equation}\label{eq:step}
\left(\hb(x,0),\ub(x,0),z(x,0)\right) = \begin{cases} \left(\displaystyle h_-,u_-,z_-\equiv\sqrt{\frac{a_-}{h_-}}\right), &x<0, 
\\
\\ 
\left(\displaystyle h_+,u_+,z_+\equiv\sqrt{\frac{a_+}{h_+}}\right), &x>0,
\end{cases}
\end{equation}
with the constraint $u_--2\mu \sqrt{h_-}=u_+-2\mu \sqrt{h_+}$.
The solution of interest is the simple wave solution
\begin{align}
  \label{eq:zp}
  &Q_{\sigma\mu}(\hb,z)=Q_{\sigma\mu}(h_-,z_-) = Q_{\sigma\mu}(h_+,z_+) ,\quad
  V_\mu = r_{-\mu}+3\mu\sqrt{\hb} = x/t.
\end{align}
Equation \eqref{eq:zp} implicitly determines $\hb$ and $z^2=a/\hb$ as
functions of $x/t$. It describes the fast RW with $h_-<h_+$. To simplify the discussion of the solution, we
restrict our study to fast solitary waves ($\sigma = +1$). In this
case, the Riemann problem for the modulation system models the
interaction of an incident solitary wave with parameter $z_-$,
initially located far to the left of the initial step $x<0$,
interacting with a RW or a DSW generated by the initial step.

It is important to note here that, although the solitary wave limit of
the SGN-Whitham equations \eqref{eq:hb}--\eqref{eq:z} do not admit DSW
modulation solutions, they can be used to determine the nature of the
interaction between a solitary wave and a DSW.  The only region of
space-time where equations \eqref{eq:hb}--\eqref{eq:z} break down is
within the DSW itself.  Outside of the DSW, equations
\eqref{eq:hb}--\eqref{eq:z} are perfectly valid, so that the simple
wave solution \eqref{eq:zp} applies \textit{outside of the DSW}.  If
all we wish to determine is whether the solitary wave has been
transmitted or trapped by the DSW and, if transmitted, the transmitted
solitary wave amplitude, we can use the Riemann invariant relation from \eqref{eq:zp} with $h_- > h_+$ to
ascertain this as described below. This concept of
\textit{hydrodynamic reciprocity} was first theoretically predicted
and simultaneously experimentally observed in the interfacial fluid
dynamics of a viscous fluid conduit \cite{maiden_solitonic_2018}.  The
concept has since been applied to solitary wave-DSW interaction for
other equations
\cite{sprenger_hydrodynamic_2018,sande_dynamic_2021,ryskamp_oblique_2021,ablowitz_solitonmean_2023}.
In order to describe the interaction of the solitary wave
\textit{within the DSW}, one must appeal to multi-phase Whitham
modulation theory as has been carried out for the KdV equation
\cite{ablowitz_solitonmean_2023}.

The first equation in~\eqref{eq:zp} yields a relation between the left
and right parameters $(h_\pm,z_\pm)$ of the initial condition, i.e., a
relation between the parameters of the incident and transmitted waves
\begin{equation}
  \label{eq:transmission}
  f_{\sigma \mu}(z_+)+ \ln h_+ =f_{\sigma \mu}\left( z_-\right) + \ln
  h_-,\quad z_\pm^2 = \frac{a_\pm}{h_\pm}. 
\end{equation}
If transmitted, the solitary wave amplitude $a_+>0$ is determined by
solving for $z_+$ in eq.~\eqref{eq:transmission}.

\begin{figure}[H]
  \centering
  (a)\includegraphics[scale=0.5]{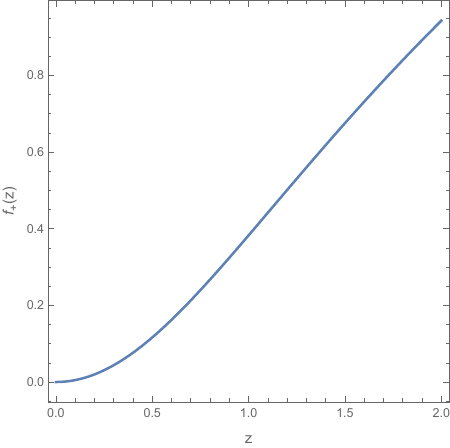}\hspace{3mm}
  (b)\includegraphics[scale=0.51]{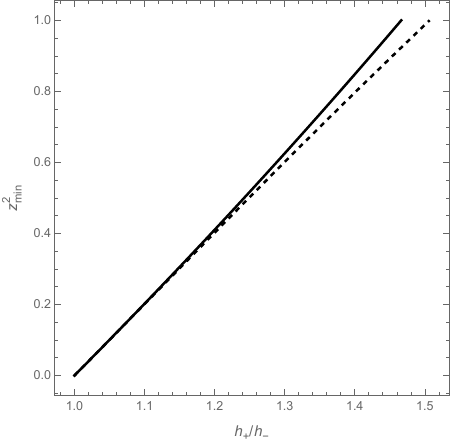}\hspace{3mm}
  (c)\includegraphics[scale=0.5]{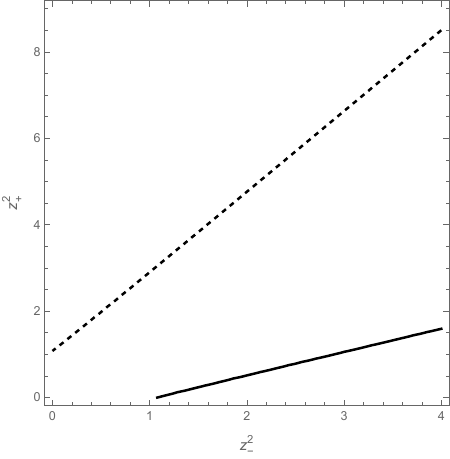}
  \caption{a) Variation of $f_+(z)$. b) The critical transmission curve $z_{\min}^2(h_+/h_-)$
   for the overtaking interaction between a fast
    solitary wave and a fast RW ($h_-<h_+$) given by the exact formula  \eqref{eq:transmission2} (continuous line) and the convex approximation formula \eqref{zmin_fitting} (dashed line). c) Relation between $z_{+}^2$ and
    $z_{-}^2$ for the interaction with a fast RW,
    $(h_-,h_+)=(1,1.5)$ (solid line) and a fast DSW,
    $(h_-,h_+)=(1.5,1)$ (dashed line).}
  \label{fig:fast}
\end{figure}

In the interaction with a fast simple wave, where $\sigma \mu=+1$, one
can choose the constant of integration in \eqref{eq:f} such that
$f_+(0)=0$. In that case, $f_+(z) \geq 0$ (see Fig.~\ref{fig:fast}(a))
and one can find $z_+$ from Eq. \eqref{eq:transmission} if
\begin{equation}
  f_+\left( z_-\right) + \ln\left( \frac{h_-}{h_+}\right) >0.
\end{equation}
This condition is called the \textit{transmission condition}. It is
always fulfilled in the interaction with a fast DSW where
$h_->h_+$. In the interaction with a fast RW, this condition can be
written in the form:
\begin{equation}
  \label{eq:transmission2}
  z_- > z_{\rm min}(h_+/h_-) = f_+^{-1} \left( \ln \left(\frac{h_+}{h_-}\right)\right),
\end{equation} 
plotted in Fig.~\ref{fig:fast}(b)---solid line. If the normalized amplitude of the
incident solitary wave is smaller than $z_{\rm min}$, then the wave
becomes trapped inside the RW, similar to \cite{maiden_solitonic_2018}
in the case of the KdV and conduit equations, see Fig.~\ref{fig:solitary-wave-mean-schematic}(b).

If the approximate  Riemann invariant $\tilde Q_+$ \eqref{eq:qfit} is used in the transmission relation \eqref{eq:transmission} then the fitting counterpart of the exact  condition \eqref{eq:transmission2} assumes the form
\begin{equation}
z_->z_{\min}(h_+/h_-)={\tilde f}_+^{-1}\left(\frac{h_+}{h_-}\right),
\label{zmin_fitting}
\end{equation}
where
\begin{equation}
{\tilde f}_+(z)=\frac{2^{2/5} 3^{21/10} \alpha^{1/2}}{(\alpha
    +1)^{2/5} (4-\alpha)^{21/10}}, \quad \alpha=\sqrt{1+z^2}.
    \end{equation}
    The curve \eqref{zmin_fitting} is plotted in
    Fig.~\ref{fig:fast}(b) in dashed line. There is a very close
    agreement between the exact and convex approximation curves.

If the fast solitary wave is placed in front of the fast DSW two scenarios are possible depending on the solitary wave amplitude $z_+$ relative to the DSW leading edge amplitude  $z^*_+$ (see Equation \eqref{eq:dsw-soli-edge} below).  If $z_+ > z^*_+$ then the solitary wave propagates faster than the DSW and no interaction occurs. If $z_+< z^*_+$ then the DSW overtakes the solitary wave and the latter gets trapped inside the DSW, see Fig.~\ref{fig:solitary-wave-mean-schematic}(d) for the illustration. The analysis of the trapping dynamics characterised by the formation of travelling breathers \cite{ablowitz_solitonmean_2023, mao_observation_2023}  is beyond the scope of this paper.

\begin{figure}[H]
  \centering
  (a)\includegraphics[scale=0.44]{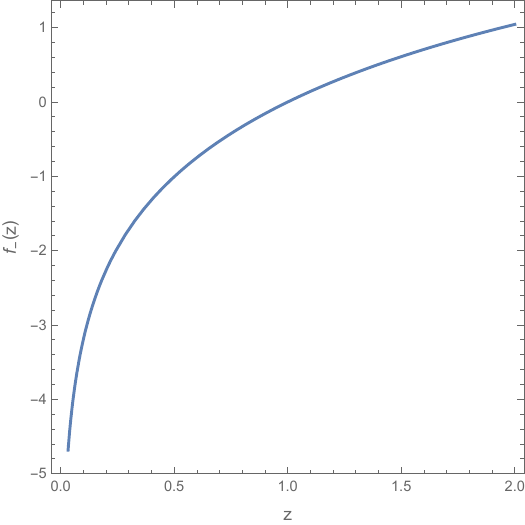}\qquad
  (b)\includegraphics[scale=0.5]{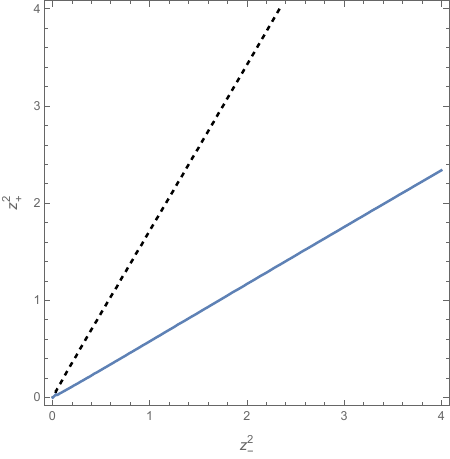}
  \caption{a) Variation of $f_-(z)$. b) Relation between $z_+^2$ and
    $z_-^2$ for the head on interaction of a fast solitary wave with a
    slow DSW, $(h_-,h_+)=(1,1.5)$ (solid line) and a slow RW,
    $(h_-,h_+)=(1.5,1)$ (dashed line). }
  \label{fig:slow}
\end{figure}

In the head on interaction with a slow simple wave, where $\sigma \mu=-1$,
$f_-(z)$ is singular at $z=0$ (see Fig.~\ref{fig:slow}(a) where
$f_-(1)=0$), and the range of $f_-(z)$ spans the whole real axis. One
can always find $z_-$ from Eq.~\eqref{eq:transmission}.  The
transmission condition is always satisfied in this case, i.e., the
incident fast solitary wave is never trapped by the RW or DSW.

\subsection{DSW: solitary wave edge}
\label{sec_DSW}
The transmission condition \eqref{eq:transmission} can also be used to approximate the SGN DSW's solitary wave leading edge amplitude and speed.  When $h_- > h_+$, a DSW is generated by the simple wave Riemann problem \eqref{eq:step}.  The DSW's solitary wave leading edge amplitude $a_+ = h_+ (z_+^*)^2$ can be estimated by evaluating the Riemann invariant at the transmission/trapping bifurcation point when $z_- = 0$. In other words,  a fast solitary wave of infinitesimally small amplitude gets transmitted exactly to the DSW leading edge. As follows from the transmission relation \eqref{eq:transmission} this corresponds to
\begin{equation}
   \label{eq:dsw-soli-edge}
   z_+^* = f_+^{-1} \left ( \ln\left ( \frac{h_-}{h_+} \right ) \right ) .
\end{equation}
The DSW solitary wave amplitude is $a_+ = h_+ (z_+^*)^2$ and edge
speed is therefore $c_{\rm s}(a_+,h_+,u_+) = u_+ + \sqrt{h_+ + a_+}$.

The SGN DSW leading solitary wave amplitude prediction from eq.~\eqref{eq:dsw-soli-edge} is an approximation of the amplitude that results from integrating the appropriate integral curve (in this case, for $h_- > h_+$, a 3-wave) of the full SGN-Whitham equations.  The approximation rests upon the hypothesis of hydrodynamic reciprocity---in which the solitary wave Riemann invariant $Q_{+1}(\hb,z)$ \eqref{eq:f} is the same behind and ahead of a DSW \cite{maiden_solitonic_2018}---and a more subtle, convexity condition that the full Whitham modulation system remains strictly hyperbolic along the entire solitary wave trajectory for all so-called admissible states; see, Section 3.2 in reference \cite{sande_dynamic_2021} for a discussion. Although these conditions are exactly true, within the context of Whitham modulation theory, for a fast (slow) solitary wave interacting with a fast (slow) rarefaction wave mean-field, they are assumed to be true for the fast (slow) DSW mean-field.  Because of this, we refer to eq.~\eqref{eq:dsw-soli-edge} as the \textit{mean-field DSW approximation}.

If instead of the exact Riemann invariant \eqref{eq:f}, its convex
(DSW fitting) approximation \eqref{eq:qfit} is used in the
transmission condition \eqref{eq:zp}, the DSW leading edge amplitude
$\tilde{a}_+ = h_+(\tilde{z}_+^*)^2$ is found by solving the equation
\cite{el_unsteady_2006}
\begin{equation}\label{leada}
  \begin{split}
    \frac{h_-/h_+}{((\tilde{z}_+^*)^2+1)^{1/4}}-\left(\frac{3}{4-\sqrt{(\tilde{z}_+^*)^2+1}}\right)^{21/10} 
     \left(\frac{2}{1+\sqrt{(\tilde{z}_+^*)^2+1}}\right)^{2/5}=0 \, .
  \end{split}
\end{equation}
Equation \eqref{leada} is subject to the admissibility  conditions---the convexity restrictions 
ensuring monotone behaviour of the DSW edge speeds as functions of $h_-, h_+$ \cite{hoefer_shock_2014}. This yields the admissible range of initial steps  $1< h_-/h_+ \lesssim 1.43$ for which the SGN DSW fitting results are applicable \cite{el_unsteady_2006}.
 \vspace{20mm}
\begin{figure}
  \centering
  \includegraphics[scale=0.4]{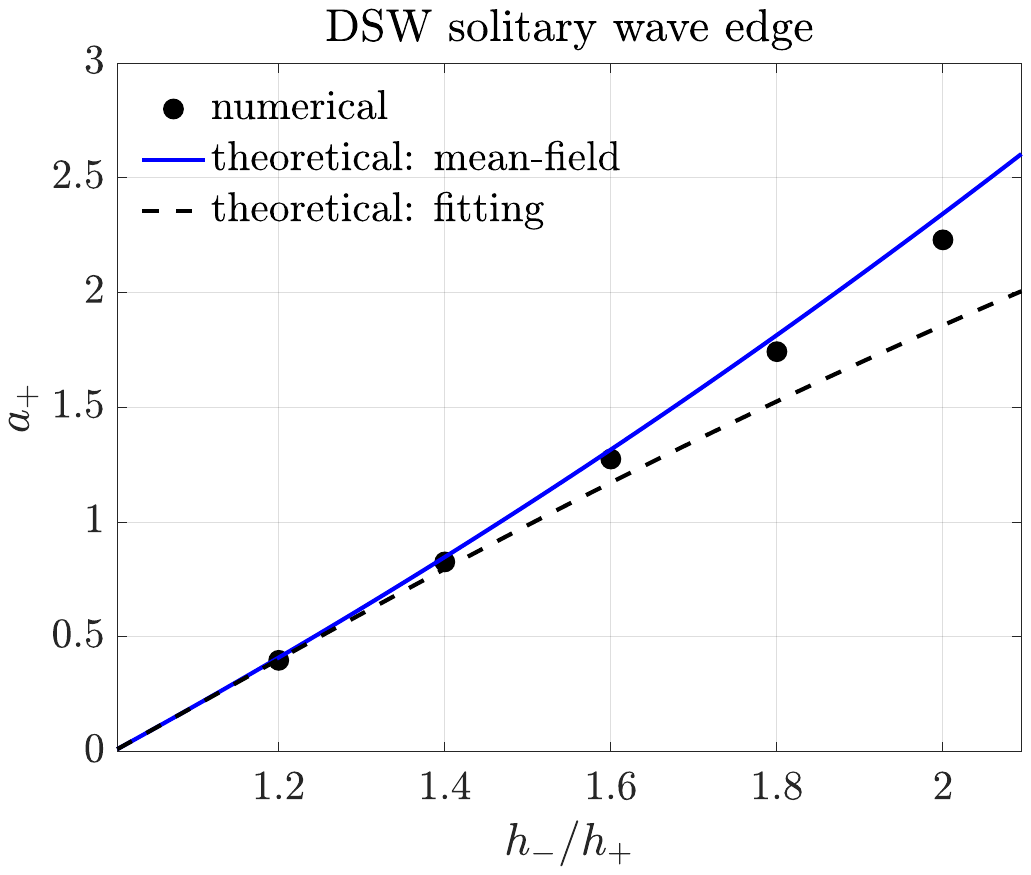}  
  \vspace{-30mm}
  \caption{DSW lead solitary wave amplitude. Solid line: mean-field DSW solution \eqref{eq:dsw-soli-edge}; dashed line: DSW fitting formula \eqref{leada}; circles: SGN numerical simulations.}    
  \label{fig:dsw_lead}
\end{figure}

The small-jump expansions of \eqref{eq:dsw-soli-edge} and
\eqref{leada} give
\begin{align}
  \label{leada_expansion_exact}
  \frac{a_+}{h_+} &= 2 \delta + \frac{1}{6}\delta^2 + \frac{127}{180} \delta^3
                    + \mathcal{O}(\delta^4), \\
  \label{leada_expansion_fitting}
  \frac{\tilde{a}_+}{h_+} &= 2 \delta + \frac{1}{6}\delta^2 -
                            \frac{71}{108} \delta^3 +
                            \mathcal{O}(\delta^4),
\end{align}
respectively, where $\delta = \frac{h_-}{h_+} - 1 \ll 1$.  To leading
order, both results agree with the famous result for the KdV equation
that the lead solitary wave amplitude is twice the initial DSW jump
height \cite{gurevich_nonstationary_1974}. The mean-field DSW and DSW fitting approximations agree at second order as well,
consistent with the agreement between the expansions of the exact
\eqref{eq:qp} and convex approximation \eqref{eq:small_amp} for the
solitary wave Riemann invariants.  The curves $a_+(h_-/h_+)$ given by
\eqref{eq:dsw-soli-edge}, \eqref{leada} along with the values of $a_+$
extracted from the numerical simulations are shown in
Fig.~\ref{fig:dsw_lead}.  One can see that both the mean-field DSW equation
\eqref{eq:dsw-soli-edge} and the DSW
fitting formula \eqref{leada} provide very good approximations of
the SGN DSW solitary wave edge amplitude within the admissible range
of initial steps.  The numerical results in Fig.~\ref{fig:dsw_lead} suggest that the mean-field DSW prediction somewhat improves upon the DSW fitting result.

\subsection{Transmission: comparison between analytical and numerical results}

\begin{figure}[H]
\begin{center}
\begin{tabular}{ccc}
\begin{minipage}{2.3in}
\hskip -0.14in
 \includegraphics[width=2.3in]
  {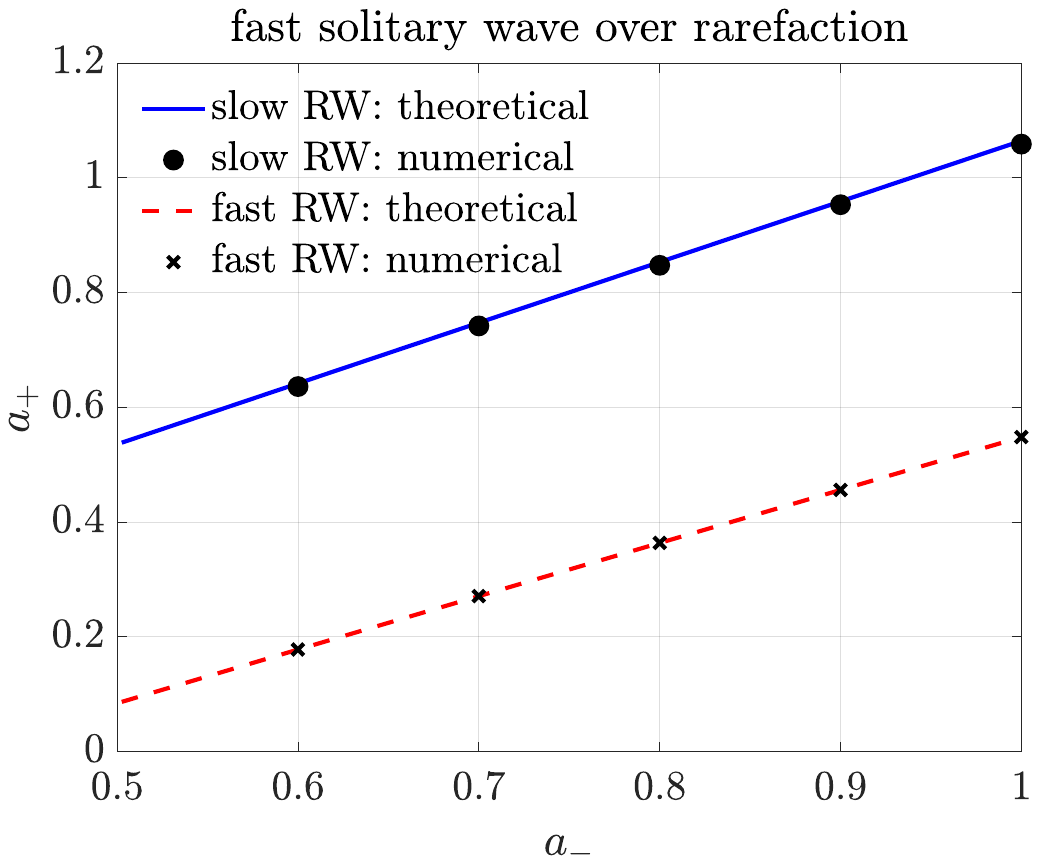}
\end{minipage} 
&
\begin{minipage}{2.3in}
 \hskip -0.66in
 \includegraphics[width=2.3in]
 {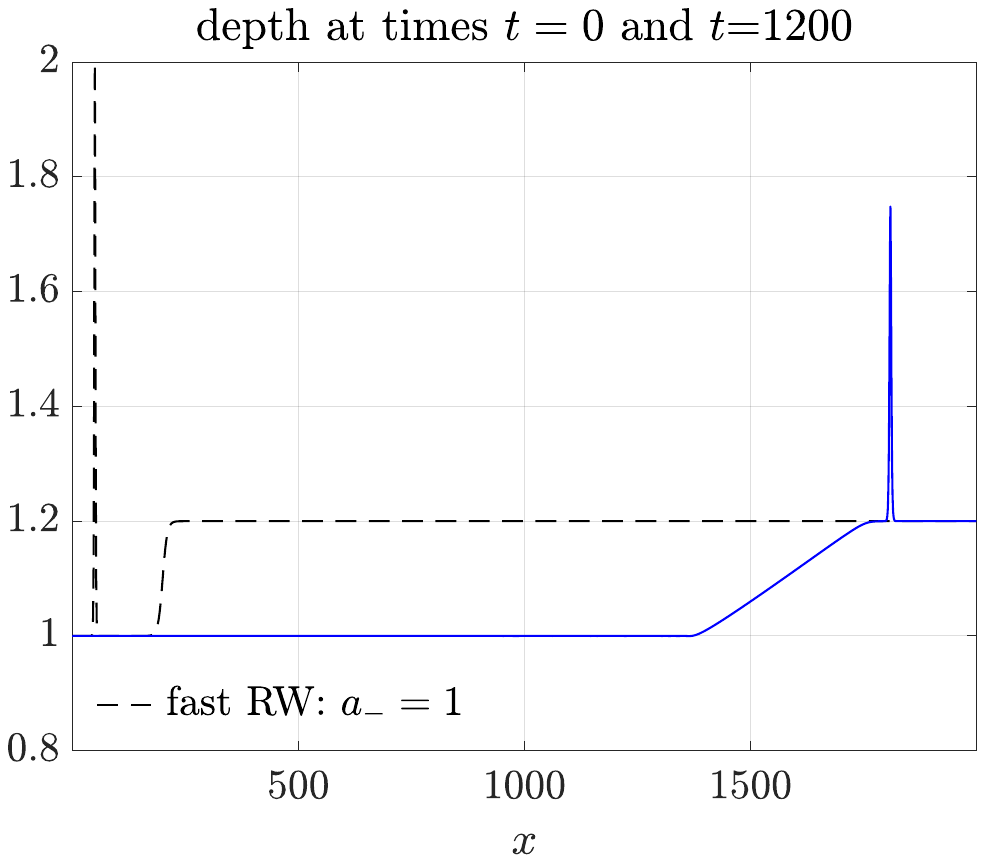}
\end{minipage}
&
\begin{minipage}{2.3in}
 \hskip -1.2in
 \includegraphics[width=2.3in]
   {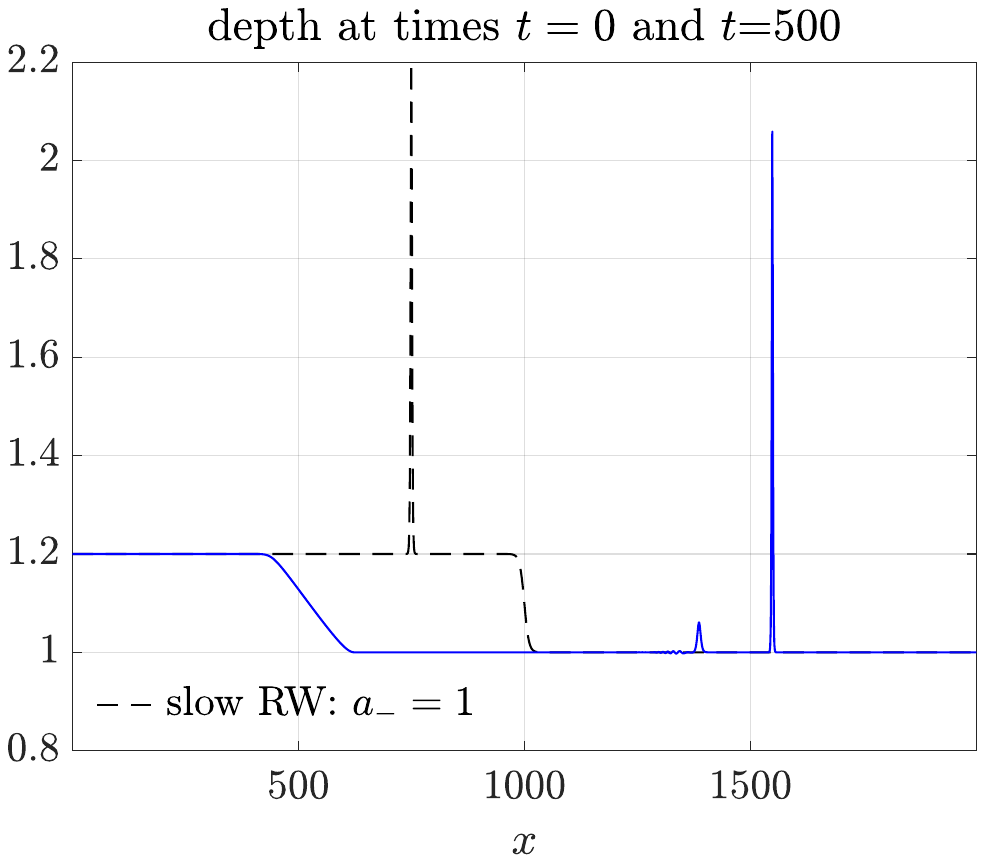}
\end{minipage}
\end{tabular}
\end{center}
\vskip -1.0in
\caption{Left figure: 
 comparison between the transmission relation ~\eqref{eq:transmission} and
   the numerical simulation for the fast solitary wave interaction with a fast RW
   (dashed red line) and   a slow rarefaction wave (solid blue line). Middle figure: an example of the
   interaction with a fast RW.
   Right figure: an example of the interaction with a slow RW.}
  \label{fig:fast_comp}
\end{figure}
\begin{figure}[H]\begin{center}
\begin{tabular}{ccc}
\begin{minipage}{2.3in}
\hskip -0.14in
 \includegraphics[width=2.3in]
  {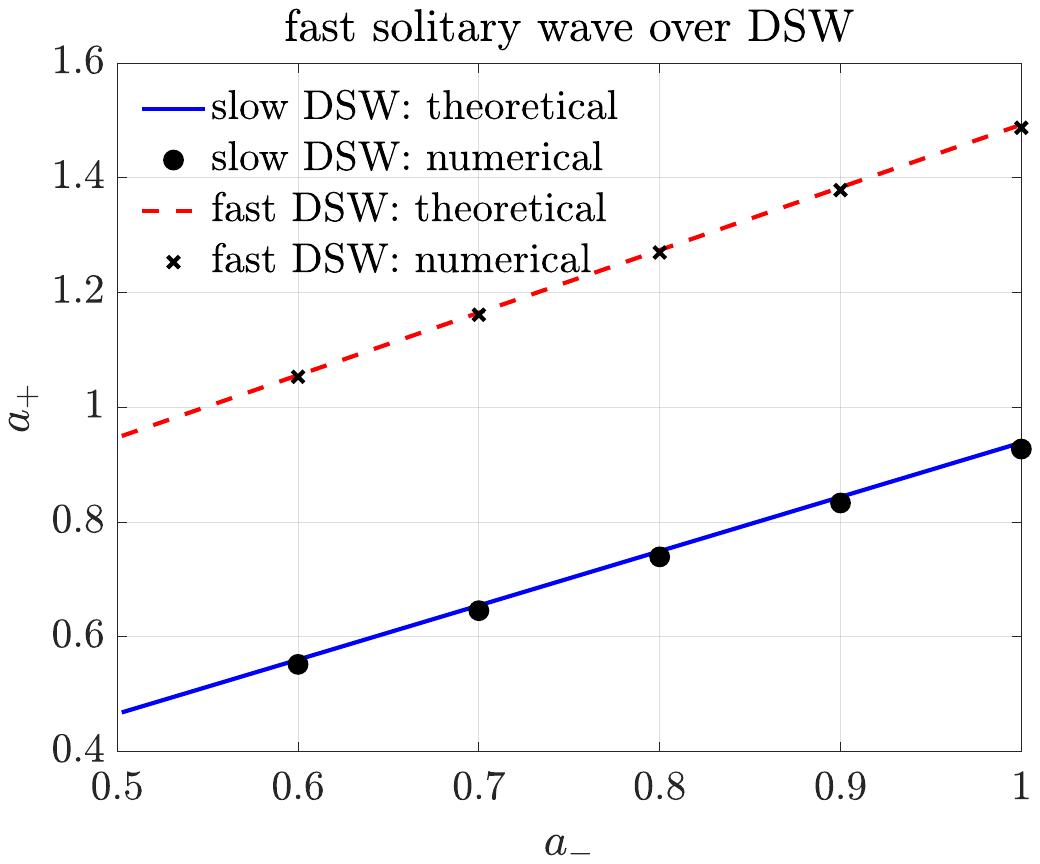}
\end{minipage}
&
\begin{minipage}{2.3in}
 \hskip -0.66in
 \includegraphics[width=2.3in]
  {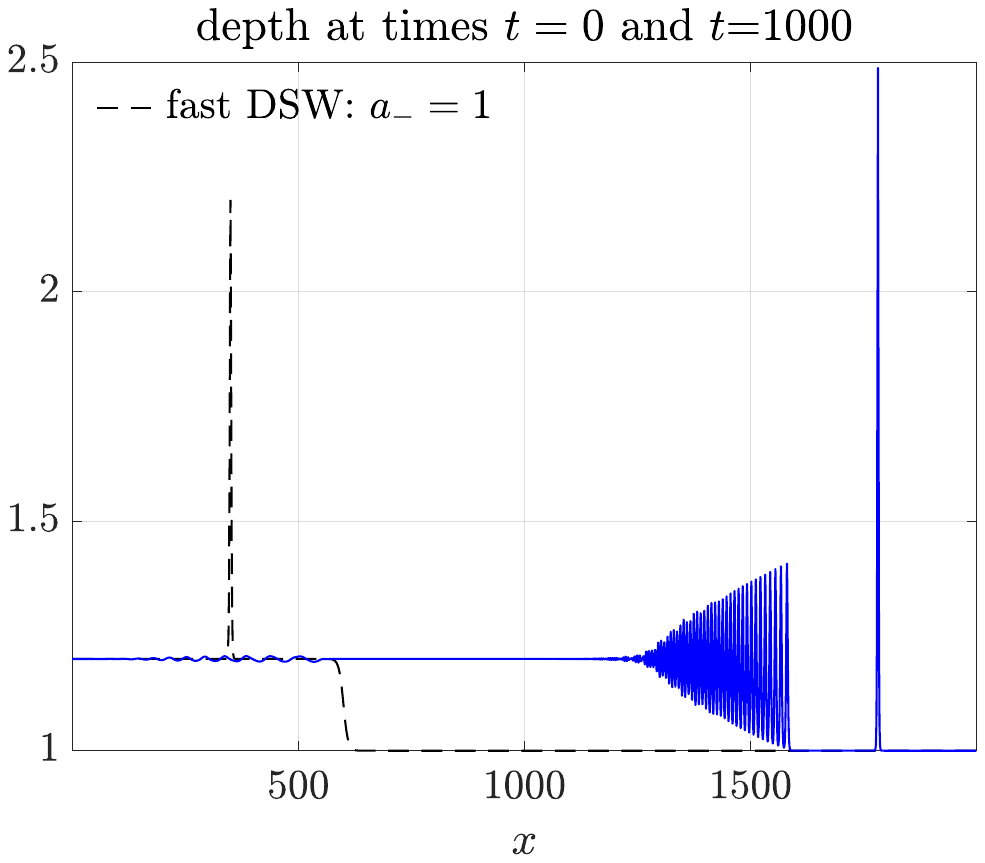}
\end{minipage}
&
\begin{minipage}{2.3in}
 \hskip -1.2in
 \includegraphics[width=2.3in]
  {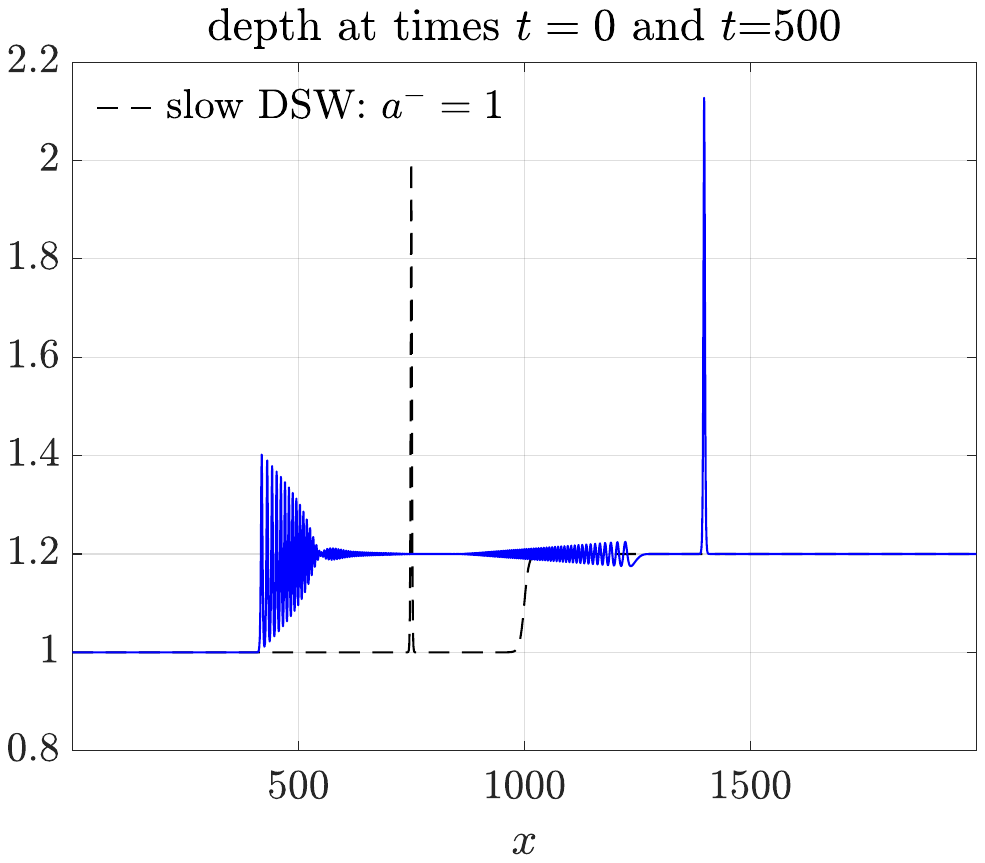}
\end{minipage}
\end{tabular}
\end{center}
  \vskip -0.9in
  \caption{Left figure: comparison between the transmission relation ~\eqref{eq:transmission} and
   the numerical simulation for the fast solitary wave interaction with  a fast DSW (dashed red line) and a slow DSW
   (solid blue line). Middle figure: an example of the interaction with a
   fast DSW.
   Right figure: an example of the interaction with a slow DSW.}  
  \label{fig:slow_comp}
\end{figure}

Figure \ref{fig:fast_comp} shows the interaction of a fast solitary
wave with fast and slow rarefaction waves. The amplitude of the
transmitted solitary wave $a_+$ obtained by the direct numerical
simulation of the SGN equations \eqref{eq:186a}, \eqref{eq:186b}
compares well with the analytical prediction
\eqref{eq:transmission}. The details of the numerical scheme are given
in Appendix \ref{sec:SGN_meth}.  In the case of a fast solitary wave
with a slow rarefaction wave, the amplitude of the transmitted
solitary wave is always larger than that of the incident solitary
wave.  Physically speaking, the slow rarefaction wave is formed by
retracting a piston to the right, so it will accelerate the
right-going solitary wave (see Figure \ref{fig:fast_comp}). A new
feature also appears in this case: a small amplitude solitary wave
forms behind the transmitted large amplitude solitary wave.  This is
in contrast with the typical radiation emitted when a solitary wave
interacts with another wave in non-integrable systems. Despite the
generation of additional solitary waves in this Riemann problem due to
the non-integrable nature of the SGN equations, the accuracy of the
analytical predictions from Whitham theory is remarkable.  \vskip
-0.8in
Figure \ref{fig:slow_comp} shows the interaction of a solitary wave
with fast and slow DSWs.  The analytical formulas are in agreement
with direct numerical simulations.  This time, small amplitude
radiation is generated after the solitary wave passes through the slow
DSW.  \vskip -0.8in

\section{Conclusion}

We have studied the interaction of a solitary wave with a slowly
varying mean flow for the Serre-Green-Naghdi (SGN) equations that
model fully nonlinear, bidirectional shallow water gravity waves over
a flat bottom.  This was achieved by the determination and analysis of
the exact solitary wave limit of the Whitham modulation equations for
the SGN system.  The derivation of the Whitham equations was performed
via the averaged conservation law approach in both Eulerian and
Lagrangian coordinates. The SGN system is not integrable so the
SGN-Whitham equations cannot be diagonalized, which makes their
analysis a nontrivial problem.

Due to the singular nature of the solitary wave limit, the appropriate
choice for the modulation parameters is crucial.  We utilize the wave
action density, which is shown to be a particularly well-behaved
conserved quantity in the solitary wave limit.  The resulting
solitonic modulation system consists of the shallow water equations
for the mean flow coupled to an amplitude equation for the solitary
wave.  Although the SGN solitonic system is not diagonalizable, its
restriction to simple waves for the mean flow equations admits Riemann
invariants that we use to analytically describe the head-on and
overtaking interactions of a solitary wave with a rarefaction wave and
dispersive shock wave (DSW).  These scenarios lead to solitary wave
trapping or transmission by the mean flow.  The analytical results are
shown to be in excellent agreement with corresponding numerical
solutions of the full SGN equations. This work extends previous
results on solitary wave-mean flow interaction in unidirectional,
non-integrable systems (BBM and conduit equations) to the
bidirectional case.

One important outcome of this paper is the comparison between the
exact analytical results from SGN modulation theory for overtaking
interactions with the results obtained from a simpler approach that is
based on the DSW fitting method. The latter method, involving several
major assumptions of the full moduation system (strict hyperbolicity,
convexity), has been successfully applied to many dispersive
hydrodynamic equations.  It yields exact results for integrable
equations, relies only upon knowledge of the linear dispersion
relation and the dispersionless, mean-flow equations, and provides
consistently good comparisons with direct numerical simulations for
non-integrable equations.  However, analytical estimates of DSW
fitting theory accuracy have not been available.  In this paper, we
show that the DSW fitting results for the SGN equation are not exact
but are accurate to the second order in the solitary wave amplitude,
beyond the first order accurate Korteweg-de Vries approximation.  This
comparison between the exact and fitting results for the SGN-Whitham
equations provides an important asymptotic benchmark for the DSW
fitting method in non-integrable equations.

This work paves the way for additional studies of solitary wave
interactions with mean flows in bidirectional, strongly nonlinear,
physically relevant systems. One of the future directions could be the
generalization of the developed theory for bidirectional solitary
wave-mean flow interactions to soliton gases---large, random ensembles
of interacting solitary waves \cite{suret_soliton_2024}.

\medskip

{\bf Acknowledgment} TC, GE, SG and MH would like to thank the Isaac
Newton Institute for Mathematical Sciences, Cambridge, for support and
hospitality during the satellite program ``Emergent phenomena in
nonlinear dispersive waves'' at Northumbria University, Newcastle, UK
where work on this paper was undertaken. This work was supported by
EPSRC grant no EP/R014604/1 and NSF grant no DMS-2306319.  KMS thanks
the Institut M\'{e}canique et Ing\'{e}nierie of Marseille and
INRIA-Sophia-Antipolis-M\'{e}diterran\'{e}e for financial support
during his stay at IUSTI, Marseille. The authors thank M.~Rodrigues
for useful discussions.

\appendix
\section{ Lagrangian and  Eulerian descriptions}
\label{Appendix_A}
If $t$ is time, and $q$ is the mass Lagrangian coordinate, the spatial
Eulerian coordinate is defined through the motion of the continuum
$x=x(t,q)$. The mass conservation equation in the Lagrangian
coordinates can be written as $\rho(t,q)x_q=\rho_0(q)$, where
$\rho(t,q)$ and $\rho_0(q)$ are the actual and the reference mass
densities, so that $\rho(t,q)dx=\rho_0 (q)dq$. One can see that if we
choose $\rho_0 (q) \equiv 1$, the Lagrangian variable $q$ will
effectively coincide with the mass (one-dimensional). A general
conservation law in Lagrangian coordinates $(t,q)$
\begin{equation*}
  a_t+b_q=0, 
\end{equation*}
is expressed in Eulerian coordinates $(t,x)$ as
\begin{equation*}
  \left({\rho a}\right)_t+\left(\rho ua+b\right)_x=0 .
\end{equation*}
When looking for traveling wave solutions that depend on
$\tilde \xi =q-\tilde c t$ in Lagrangian coordinates, or $\xi =x-ct$
in Eulerian coordinates, the velocities $\tilde c$ and $c$ are related
by the mass conservation law
\begin{equation*}
  \rho(u-c)=-\tilde c. 
\end{equation*} 
We will first pass to the solitary wave limit in the modulation
equations expressed in terms of mass Lagrangian coordinates and then
we transform the limit equations to Eulerian form.

The governing Euler-Lagrange equations derived from the Lagrangian
\eqref{basic_lagrangian} are
\begin{equation}
	\tau_{t}- u_{q}=0,\quad u_{t}+p_{q}=0, 
	\label{basic_dispersion}
\end{equation}
where $p$ is given in \eqref{definition_pressure_0}.
Here, $\tau = 1/h$ is the specific volume, $u$ is the velocity, and
$\tilde e(\tau, \tau_t)$ is the potential
\eqref{SGN_energy_mass_coordinates}.  The system admits the energy
equation \eqref{energy_Lagrangian_0} and Bernoulli equation
\eqref{Bernoulli_Lagrangian_0}.
Following Whitham \cite{whitham_linear_1999}, we are looking for the
solution of (\ref{basic_dispersion}) subject to
\eqref{definition_pressure_0} in the form
\begin{equation*}
  \dbinom{\tau}{u}=\dbinom{\tau(T,X,\theta,\epsilon)}{u(T,X,\theta,
    \epsilon)}, \quad T=\epsilon t,\quad X=\epsilon q, \quad
  \theta=\frac{\Theta\left(T,X\right)}{\epsilon}, 
\end{equation*}
where $0 < \epsilon \ll 1$ is a small, positive parameter. The
function $\Theta(T,X)$ is called the phase. We define the local wave
number $k$ and the local frequency $\omega$ by the relations:
\begin{equation}
  k=\Theta_X, \quad \omega=-\Theta_T.
  \label{definition_wave_number}
\end{equation}
We will suppose that the solution is $2\pi$-periodic with respect to
the rapid variable $\theta$.  Below, for any function
$f(T,X,\theta,\epsilon)$, $\bar f$ is the $2\pi$-period-average with
respect to $\theta$
\begin{equation}
  \bar f
  (T,X,\epsilon)=\frac{1}{2\pi}
  \int_0^{2\pi}f(T,X,\theta,\epsilon)\,d\theta.  
  \label{lagrangian_averaging}
\end{equation}
Expanding $f$ in an asymptotic series in $\epsilon$
\begin{equation*}
	f=f_0(T,X,\theta)+\epsilon f_1(T,X,\theta)+...,
\end{equation*}	
and substituting this ansatz into the governing system
\eqref{basic_dispersion}, we obtain a system of ordinary differential
equations with respect to $\theta$ at leading order. One has the
following first integrals (the zero subscript is omitted)
\begin{equation}
  \tilde{c}\tau +u= \tilde{c}\overline \tau+\overline u, \quad
  -\tilde{c}u+p=-\tilde{c}\overline u +\overline p, \quad
  \tilde{c}=\frac{\omega}{k}. 
  \label{stationary_solutions}
\end{equation}
The system (\ref{stationary_solutions}) admits the following useful
consequences
\begin{align}
  \overline{u^2}-(\overline u)^2=\tilde{c}^2(\overline
  {\tau^2}-(\overline \tau)^2), 
  \label{1}
  \\
  \overline{u\tau}-{\overline u} \; {\overline \tau}
  =-\tilde{c}(\overline {\tau^2}-(\overline \tau)^2), 
  \label{2}
  \\
  \overline{p u}- {\overline p}\;  {\overline u}
  =\tilde{c}^3(\overline {\tau^2}-(\overline \tau)^2), 
  \label{3}
  \\
  \overline{p \tau}- {\overline p} \;  {\overline \tau}
  =-\tilde{c}^2(\overline {\tau^2}-(\overline \tau)^2). 
  \label{4}
\end{align}
To leading order, $\mathcal{O}(1)$ as $\epsilon \to 0$, one has
\begin{equation}
  \tilde e (\tau,\tau_t)\approx \tilde e (\tau, \eta), \quad
  \eta=-\omega\tau_\theta, \quad  
  p=-\left(\tilde e_\tau +\omega\left(\tilde e_{\eta}\right)_\theta \right). 
\end{equation}
The system (\ref{stationary_solutions}) also admits the first integral
in the form:
\begin{equation}
  -\tilde{c}\left(\tilde e-\eta \tilde e_\eta
    +\frac{u^2}{2}\right)+pu=-\tilde{c}\left(\overline{ \tilde e-\eta
      \tilde e_\eta  +\frac{u^2}{2}}\right) +\overline{p u}.
  \label{stationary equation} 
\end{equation}
The integral (\ref{stationary equation}) allows us to obtain the
nonlinear dispersion relation because the solution is
$2\pi$-periodic. At $\mathcal{O}(\epsilon)$, after averaging with
respect to the rapid variable $\theta$, one obtains the following
systems of five compatible conservation laws containing only the
leading order terms (the zero subscript is again omitted):
\begin{align*}
  {\overline \tau}_T-{\overline u}_X=0, \\
  {\overline u}_T+{\overline p}_X=0, \\
  \left(\overline{ \frac{u^2}{2}+\varepsilon}\right)_T+\left(\overline{ pu}\right)_X=0,\quad \varepsilon=\tilde e-\eta \tilde e_\eta, \\
  \left(\overline{\tau u-k\tau_\theta\frac{\partial \tilde
  e}{\partial\eta }}\right)_T-\left(\overline{\frac{u^2}{2}-\tau
  p-\tilde e}\right)_X =0,  \\
	k_T+\omega_X =0.
\end{align*}
The last equation is just the compatibility condition coming from the
definition of the phase (\ref{definition_wave_number}). It can be also
written in the form
\begin{equation*}
  k_T+\left(\tilde{c} k\right)_X =0, \quad \tilde{c}=\frac{\omega}{k}. 
\end{equation*}
Introducing
\begin{equation*}
  \Delta =\overline{u\tau}-{\overline u}\; {\overline \tau}, \;
  \overline{\varepsilon}=\overline{\tilde e-\eta \tilde e_\eta}, \;
  E=\overline{\varepsilon}+\frac{\overline{u^2}-\overline{u}^2}{2}, \;
  \Sigma=\overline{\tau_\theta \tilde e_\eta}, 
\end{equation*}
one can obtain a Gibbs-type identity relating the unknowns (see
\cite{gavrilyuk_large_1994,gavrilyuk_model_1995} for proof)
\begin{equation}
  dE+\overline{p} \; d\overline{\tau} +\tilde{c}d\Delta=\omega d\Sigma.
  \label{Gibbs_identity} 
\end{equation}
It can also be written in the form:
\begin{equation}
	d\overline{\varepsilon}+\overline{p} \; d\overline{\tau} -\frac{\tilde{c}^2}{2}d\delta=\omega d\Sigma,\quad \delta=\overline {\tau^2}-(\overline \tau)^2.
	\label{Gibbs_identity_new} 
\end{equation}
The Gibbs identity \eqref{Gibbs_identity} or
\eqref{Gibbs_identity_new} is equivalent to an algebraic nonlinear
dispersion relation coming from \eqref{stationary equation} when we
are looking for $2\pi$-periodic solutions \cite{whitham_linear_1999}.

Using the expressions for correlations (\ref{1})--(\ref{4}), the
modulation equations take the conservative form
\begin{equation}
  \label{Whitham_system}
  \begin{split}
    {\overline \tau}_T-{\overline u}_X=0, \\
    {\overline u}_T+{\overline p}_X=0, \\
    \left(\overline{\varepsilon}
    +\frac{\overline{u}^2}{2}+\frac{\tilde{c}^2}{2}\left(\overline{\tau^2}-
    \overline{\tau}^2\right)\right)_T+\left(\overline{ p}\; \overline{u}
    +\tilde{c}^3
    \left(\overline{\tau^2}-\overline{\tau}^2\right)\right)_X=0,  \\
    \left(\overline{\tau}\; \overline{u} -\tilde{c}
    \left(\overline{\tau^2} -\overline{\tau}^2\right) -k\Sigma\right)_T
    -\left(\frac{\overline{u}^2}{2} +\frac{\tilde{c}^2}{2}
    \left(\overline{\tau^2} -\overline{\tau}^2\right)-
    \overline{\tau}\;\overline{p}+ \tilde{c}^2 \left(\overline{\tau^2}
    -\overline{\tau}^2\right) -\overline{\varepsilon}+
    \omega\Sigma\right)_X =0, \\
	k_T+(\tilde{c}k)_X =0. 
  \end{split}
\end{equation}
The five conservation laws \eqref{Whitham_system} form a system of
modulation equations (the SGN-Whitham system) for the four unknowns
$k,\; \tilde{c},\; \overline{\tau}, \; \overline{u}$.  The equations
are compatible because the averaged Bernoulli law is a consequence of
the mass, momentum, energy and wave conservation laws.  This can be
proved by direct calculations. Using the Gibbs relation
(\ref{Gibbs_identity}) or (\ref{Gibbs_identity_new}), one can derive
the sixth conservation law for $\Sigma$ (an averaged entropy
conservation law)
\begin{equation}
  \left(\Sigma +\frac{\tilde{c}\delta}{k}\right)_T
  +\left(\frac{\tilde{c}^2\delta}{k}\right)_X = 0. 
  \label{new_entropy}
\end{equation}
For the proof, one can use the energy conservation law in the form
\begin{equation*}
  \left(\overline{\varepsilon} + \frac{\overline{u}^2}{2}
    +\frac{\tilde{c}^2}{2}\delta\right)_T +\left(\overline{ p}\;
    \overline{u} +\tilde{c}^3\delta\right)_X=0. 
\end{equation*}
Expanding it and using Gibbs relation in the form
(\ref{Gibbs_identity_new}), one obtains
\begin{equation*}
  -\overline{p} \;{\overline{\tau}}_T +\frac{\tilde{c}^2}{2}\delta_T
  +\omega\Sigma_T +{\overline{u}}\;{\overline{u}}_T +
  \frac{\tilde{c}^2}{2}\delta_T +\tilde{c}\tilde{c}_T
  \delta+\overline{p}_X \overline{u}+\overline{p}\; \overline{u}_X
  +\tilde{c}(\tilde{c}^2\delta)_X +\tilde{c}^2\tilde{c}_X\delta=0.
\end{equation*}
Using the averaged mass and momentum
equations, one obtains
\begin{equation*}
  \tilde{c}^2\delta_T +\omega\Sigma_T +\tilde{c}\tilde{c}_T\delta
  +\tilde{c}(\tilde{c}^2\delta)_X +\tilde{c}^2\delta \tilde{c}_X=0.
\end{equation*}
Or, dividing by $\tilde{c}$
\begin{equation*}
  (\tilde{c}\delta)_T +k\Sigma_T +(\tilde{c}^2\delta)_X
  +\tilde{c}\delta \tilde{c}_X=0. 
\end{equation*}
Dividing by $k$, one obtains, after some algebra, the conservation law
(\ref{new_entropy}).

Finally, the full, compatible SGN-Whitham system
\eqref{Whitham_system}--\eqref{new_entropy} consists of the averaged
conservation laws of mass, momentum, energy, wave conservation, the
averaged Bernoulli conservation law and that for the averaged entropy
$\Sigma$.  The last conservation law is also called the wave action
conservation law \cite{whitham_linear_1999}.

\subsection{Solitary wave limit}

The solitary wave limit is $k\rightarrow 0$, which implies
$\omega\rightarrow 0$. In this case, the modulation equations become
purely hydrodynamic. For example, the mass and momentum equation
become
\begin{equation*}
	{\overline \tau}_T-{\overline u}_X=0, \quad
	{\overline u}_T+{\overline p}_X=0,
\end{equation*}
with 
\begin{equation}
	\overline p=-\tilde e_\tau (\overline \tau, 0). 
\end{equation}
The conservation of waves equation identically vanishes because both
$k$ and $\omega$ are zero in this limit. However, in such a limit, the
equation for the wave action is non-trivial. Let us first calculate
$\Sigma$ for the full (non-averaged) SGN equations.  With
$\displaystyle \tau = \frac{1}{h}$, to leading order $\mathcal{O}(1)$,
one has
\begin{equation}
  \tilde e=\frac{h}{2} -\frac{h_t^2}{6} =\frac{1}{2\tau}
  -\frac{\tau_t^2}{6\tau^4} \sim  
  \frac{1}{2\tau}-\frac{\eta^2}{6\tau^4}, \quad \eta =-\omega\tau_\theta. 
\end{equation}
Then,
\begin{equation}
	\tilde e_\eta \sim -\frac{\eta}{3\tau^4},
\end{equation}
and 
\begin{equation}
  \Sigma = -\overline{\frac{\tau_{\theta}
      \eta}{3\tau^4}}= \frac{\omega}{3}
  \overline{\frac{\tau_{\theta}^2}{\tau^4}} 
  =    \frac{\omega}{3}\;\overline{h_{\theta}^2}. 
\end{equation}
Setting $\zeta=q-\tilde c t$ to be the traveling wave coordinate, then
\begin{equation}
  \frac{d }{d\zeta}= k \frac{d }{d\theta}.
\end{equation}
Using this, we reduce the problem to the following one.  Find the
limits
\begin{equation*}
  \lim_{k\rightarrow 0}\tilde c \displaystyle
  \left(\frac{\overline{\left(\frac{dh}{d\zeta}\right)^2}}{3k}
    +\frac{\delta}{k}\right) \quad{\rm and} \quad   \lim_{k\rightarrow
    0}{\tilde c}^2\frac{\delta}{k}. 
\end{equation*}
Since the limit of $\tilde c$ is
$\tilde c^2=h_1h_2h_3\rightarrow h_1^2h_3$ as $k \to 0$, we need only
to find
\begin{equation*}
  \lim_{k\rightarrow 0} \displaystyle
  \left(\frac{\overline{\left(\frac{dh}{d\zeta}\right)^2}}{3k}
    +\frac{\delta}{k}\right)  \quad{\rm and} \quad  \lim_{k\rightarrow
    0}\frac{\delta}{k}. 
\end{equation*}

\subsection{Computation of average values in mass Lagrangian
  coordinates}
The wavelength in mass Lagrangian coordinates is
\begin{equation}
  L=\frac{2\vert\tilde
    c\vert}{\sqrt{3}}\int_{h_2}^{h_3}\frac{hdh}{\sqrt{P(h)}}. 
\end{equation}
Also,
\begin{equation}
  \overline{\left(\frac{dh}{d\zeta}\right)^2} =
  \frac{2}{L}\sqrt{\frac{3}{\tilde
      c^2}}\int_{h_2}^{h_3}\frac{\sqrt{P}dh}{h}. 
\end{equation} 
Then, with $k=2\pi/L$, the limit as $m\rightarrow 1$ of the ratio
\begin{align*}
  \frac{\overline{\left(\frac{dh}{d\zeta}\right)^2}}{3k}
  =\frac{1}{2\pi}\frac{2}{3}\sqrt{\frac{3}{\tilde  c^2}}
  \int_{h_2}^{h_3} \frac{\sqrt{P}dh}{h} &\rightarrow
  \frac{1}{2\pi}\frac{4}{\sqrt{3}} 
  \left(\frac{(3-2n)\sqrt{n}}{3(1-n)}-{\rm ln}
    \left(\frac{1+\sqrt{n}}{\sqrt{1-n}}\right)\right)  \\
  &=	 \frac{1}{2\pi}\frac{4}{\sqrt{3}}
  \left(\frac{(3-2n)\sqrt{n}}{3(1-n)}-\frac{1}{2}{\rm
  ln}\left(\frac{1+\sqrt{n}}{1-\sqrt{n}}\right)\right) \\
  &=	 \frac{1}{2\pi}\frac{4}{\sqrt{3}}
  \left(\frac{(3-2n)\sqrt{n}}{3(1-n)}+\frac{1}{2}{\rm
  ln}\left(\frac{1-\sqrt{n}}{1+\sqrt{n}}\right)\right) .
\end{align*}
Recall,
\begin{equation}
	m=\frac{h_3-h_2}{h_3-h_1}, \quad n=\frac{h_3-h_2}{h_3}, \quad 0<n<m<1.
\end{equation}

Indeed, making the change of variables $h=h_2+t(h_3-h_2)$,
$t\in [0,1]$ and then $t=\cos^2\theta$, $\theta\in [0,\pi/2]$, one
obtains
\begin{align*}
  \int_{h_2}^{h_3} \frac{\sqrt{P}dh}{h}
  &=\frac{(h_3-h_2)^2\sqrt{h_3-h_1}}{h_3}\int_0^1 
    \frac{\sqrt{t(1-t)(1-m(1-t))}}{1-n(1-t)}dt \\
  &= \frac{2(h_3-h_2)^2\sqrt{h_3-h_1}}{h_3} \int_{0}^{\pi/2}
    \frac{\sin^2\theta \cos^2\theta
    \sqrt{1-m\sin^2\theta}}{1-n\sin^2\theta}d\theta \\
  &\rightarrow \frac{2(h_3-h_1)^{5/2}}{h_3}
    \int_{0}^{\pi/2}\frac{\sin^2\theta \cos^2\theta
    \sqrt{1-\sin^2\theta}}{1-n\sin^2\theta}d\theta, \quad \mathrm{as}
    ~ m \to 1 \\
  &= \frac{2(h_3-h_1)^{5/2}}{h_3} \left(\frac{3-2n}{3n^2}
    -\frac{1-n}{n^{5/2}}{\rm \ln} \left(\frac{1+\sqrt{n}}{\sqrt{1-n}}
    \right)\right). 
\end{align*}
The next step is the expression for $\delta/k$. One has
\begin{equation}
  \frac{\delta}{k}=	\frac{\overline{\tau^2}
    -\left(\overline{\tau}\right)^2}{k}
  =\frac{1}{2\pi}\frac{2\vert\tilde c\vert}{\sqrt{3}}
  \int_{h_2}^{h_3}\frac{hdh}{\sqrt{P(h)}}	\left( 
    \frac{\int_{h_2}^{h_3} 
      \frac{dh}{h\sqrt{P}}}{\int_{h_2}^{h_3}\frac{hdh}{\sqrt{P}}}
    -\left(\frac{\int_{h_2}^{h_3}
        \frac{dh}{\sqrt{P}}}{\int_{h_2}^{h_3}\frac{hdh}{\sqrt{P}}}\right)^2 
  \right).
\end{equation}
The expressions of the integrals 
\begin{equation}
	\int_{h_2}^{h_3}\frac{dh}{\sqrt{P}}, \quad 
	\int_{h_2}^{h_3}\frac{dh}{h\sqrt{P}}, \quad\int_{h_2}^{h_3}\frac{hdh}{\sqrt{P}}
\end{equation}
are given in \cite{tkachenko_hyperbolicity_2020}. They give us
\begin{equation}
  \frac{\delta}{k} = \frac{1}{2\pi} \frac{4\vert\tilde
    c\vert}{\sqrt{3}} \frac{1}{h_3\sqrt{h_3-h_1}}
  \left(\Pi(n,m)-\frac{\K^2(m)}{(1-n/m)\K(m)+(n/m) \E(m)}\right). 
\end{equation}
As usual, $\K(m)$ is the complete elliptic integral of the first kind
\begin{equation}
  \K(m)=\int_0^{\pi/2}\frac{d\theta}{\sqrt{1-m\sin^2{\theta}}},
\end{equation}
$\E(m)$ is the complete elliptic integral of the second kind
\begin{equation}
  \E(m)=\int_0^{\pi/2}\sqrt{1-m\sin^2{\theta}}\, d\theta,
\end{equation} 
and $\Pi(n,m)$ is the complete elliptic integral of the third kind
\begin{equation}
  \Pi(n,m) = \int_0^{\pi/2}
  \frac{d\theta}{(1-n\sin^2\theta)\sqrt{1-m\sin^2{\theta}}}. 
\end{equation}
The limit
\begin{equation}
  \lim_{m\rightarrow
    1}\left(\Pi(n,m)-\frac{\K^2(m)}{(1-n/m)\K(m)+(n/m) \E(m)}\right) 
\end{equation}
is singular. It can be obtained in the form 
\begin{equation}
  \lim_{m\rightarrow 1} \left(\Pi(n,m) -\frac{\K^2(m)}{(1-n/m)\K(m)+
      (n/m) \E(m)}\right) =\frac{n}{(1-n)^2}
  +\frac{\sqrt{n}}{2(1-n)}{\ln}
  \left(\frac{1-\sqrt{n}}{1+\sqrt{n}}\right).
\end{equation}
Hence,
\begin{equation}
  \lim_{m\rightarrow 1}\frac{\delta}{k}=\lim_{m\rightarrow 1}
  \frac{\overline{\tau^2} -\left(\overline{\tau}\right)^2}{k}
  =\frac{1}{2\pi} \frac{4}{\sqrt{3}}\left(\frac{ \sqrt{n}}{1-n}
    +\frac{1}{2}{\rm ln}
    \left(\frac{1-\sqrt{n}}{1+\sqrt{n}}\right)\right). 
\end{equation}
In particular, this implies
\begin{equation}
  \lim_{k\rightarrow 0} \displaystyle\left(
    \frac{\overline{\left(\frac{dh}{d\zeta}\right)^2}}{3k}
    +\frac{\delta}{k}\right) =\frac{4\sqrt{3}}{2\pi}
  \left(\frac{(6-2n)\sqrt{n}}{3(1-n)}+{\rm
      ln}\left(\frac{1-\sqrt{n}}{1+\sqrt{n}}\right)\right). 
\end{equation}
Finally, the limiting SGN-Whitham modulation equation for the solitary
wave field in mass Lagrangian coordinates is
eq.~\eqref{eq:sol_limit_amplitude_lagrangian} with \eqref{eq:FF} and
\eqref{eq:GG}.
Equation \eqref{eq:sol_limit_amplitude_lagrangian} is the solitary
wave limit of eq.~\eqref{new_entropy} and expresses the conservation
of wave action.  In the limit $m\rightarrow 1$, the quantity $n$
becomes eq.~\eqref{eq:n_soli}.

\section{Solitary wave limit of the modulation equations in Eulerian
  coordinates}
\label{Appendix B}

The solitary wave limit \eqref{eq:sol_limit_amplitude_lagrangian} can
also be obtained from the modulation equations written in Eulerian
coordinates \eqref{serre_gen_mod}, obtained previously
in~\cite{el_unsteady_2006,tkachenko_hyperbolicity_2020}. Here, we use
the formulation of \cite{tkachenko_hyperbolicity_2020} with the
notation $c$ for the phase velocity and $g=1$. Note that in
\cite{tkachenko_hyperbolicity_2020}, the phase velocity was denoted by
$D$ and the roots of the polynomial were $h_0<h_1<h_2$ instead of
$h_1<h_2<h_3$ used in this work.  The SGN-Whitham modulation equations
in Eulerian coordinates are
\begin{align}
	&\overline{h}_t+(p+\overline{h} c)_x=0 \\
	&(p+\overline{h} c)_t+\left(\overline{h} c^{2}+\frac{1}{2}  I_{2}+2 p c\right)_x=0 \\
	&\left(\frac{1}{2} \overline{h} c^{2}+\frac{1}{2}  I_{1} \overline{h}-\frac{1}{2}  I_{2}+ I_{3} \overline{h^{-1}}+p c\right)_t \\
	&\quad+\left(\frac{1}{2} \overline{h} c^{3}+\frac{1}{2} I_{1} \overline{h} c+
	I_{3} \overline{h^{-1}} c+\frac{3}{2} p c^{2}+\frac{1}{2} p  I_{1}\right)_x=0 ,\\
	&k_t+(k c)_x=0,
\end{align}
with
\begin{align}
  &I_1=h_1+h_2+h_3, \quad
    I_2=h_1 h_2+h_2 h_3+h_1 h_3, \quad
    I_3=h_1 h_2 h_3,\quad
	p = -\sigma \sqrt{I_3},\\
  &\overline{h}=h_1+\left(h_3-h_1\right) \frac{\E(m)}{\K(m)}, \quad
    \overline{h^{-1}} =\frac{\Pi(n, m)}{h_3 \K(m)},  \quad
    \frac{2\pi}{k}= L=4 \sqrt{\frac{h_1 h_2 h_3}{3}}
    \frac{\K(m)}{\sqrt{h_3-h_1}}, \quad n=1-\frac{h_2}{h_3}.
\end{align}
We first rewrite the quasi-linear system above in the new variables
\begin{equation}
	(h_1,h_2,h_3,c) \quad \longrightarrow \quad
	\boldsymbol{h} = \left(h_1,h_2,a=h_3-h_2,u_1=c-\sigma\sqrt{h_1+a} \right),
\end{equation}
using a symbolic computation package, e.g., Mathematica.  Note that
$h_1$, $u_1$ are the spatial averages $\overline{h}$, $\overline{u}$
in the solitary wave limit $h_2 \to h_1$. The expressions are
particularly long and not written here for brevity.  We denote the
obtained system in the abstract form
\begin{equation}
	\label{eq:M}
	\boldsymbol{h}_t + M(\boldsymbol{h}) \boldsymbol{h}_x=0,
\end{equation}
where $M(\boldsymbol{h})$ is a $4\times 4$ matrix.

The variables $h_1,h_2,a,u_1$ remain finite in the solitary wave limit
$h_2 \to h_1$ (or $ m \to 1$), and the limit of the
system~\eqref{eq:M} thus yields non-trivial modulation equations for
$h_1$, $u_1$ and $a$. The coefficients of $M(\boldsymbol{h})$ are
rational functions of $h_1$, $h_2$, $a$, $u_1$, $\E(m)$, $\K(m)$ and
$\Pi(n,m)$.  Using the series expansions
\begin{align}
	&\E(m) = 1+{\cal O}(1-m),\quad \K(m) = 2\ln(2)-\frac{\ln(1-m)}{2} +
	{\cal O}(1-m),\\
	&\Pi(n,m) = \frac{\sqrt{n}\, \tanh^{-1}(\sqrt{n})-2\ln(2)+\ln(1-m)/2}{n-1}+{\cal O}(1-m),
\end{align}
where
\begin{equation}
	\tanh^{-1}(\sqrt{n}) = \frac{1}{2}\ln\left(\frac{1+\sqrt
		n}{1-\sqrt n} \right)=\sinh^{-1}(z),\quad n = \frac{a}{h_1+a},\quad z=\sqrt{\frac{a}{h_1}},
\end{equation}
we obtain the solitary wave limit in Eulerian coordinates
\begin{align}
	&h_{1,t}+(u_1h_1)_x=0, 
		\label{eq:h}\\
	&u_{1,t}+u_1u_{1,x}+h_{1,x}=0,
		\label{eq:u}\\
	&a_t +c a_x-\sigma \,\frac{(2
		n-3) \sqrt{n}+(3-n) (1-n) \tanh
		^{-1}\left(\sqrt{n}\right)}{ (1-n)^{3/2}\left(2 \sqrt{n}-(1-n) \tanh
		^{-1}\left(\sqrt{n}\right)\right)} \, \sqrt{h_1}\,  h_{1,x} 	\label{eq:am}\\
	&\hspace{1.5cm}-\frac{3 \sqrt{n}-(3-n) \tanh
		^{-1}\left(\sqrt{n}\right)}{2 \sqrt{n}-(1-n) \tanh
		^{-1}\left(\sqrt{n}\right)}\, h_1 \, u_{1,x}=0,\nonumber\\
	&c=u_1+\sigma \sqrt{h_1+a}.
	\label{eq:c}
\end{align}
Equation \eqref{eq:am} for the wave amplitude $a$ is equivalent to the
equation for the wave action \eqref{eq:sol_limit_amplitude_lagrangian}
written in Eulerian coordinates as
\begin{equation}
  (h_1 F(n,h_1))_t+\left(h_1 u_1 F(n,h_1) + G(n,h_1)\right)_x=0,
  \label{eq:sol_limit_Eulerian} 
\end{equation} 
with $F$ and $G$ given by \eqref{eq:FF} and \eqref{eq:GG},
respectively.  Using the notation $h_1=\overline{h}$,
$u_1=\overline{u}$, and replacing the amplitude $a$ by the
dimensionless amplitude
$\displaystyle z=\sqrt{\frac{a}{\overline{h}}}$, we recover from
\eqref{eq:h}--\eqref{eq:am} equations \eqref{eq:hb}--\eqref{eq:z}. 

\section{Compatibility condition}
\label{Appendix C}
{\bf Theorem 1}

Consider an overdetermined system of equations for the unknown
$w(x_1,x_2,x_3)$
\begin{equation}
  w_{x_1}=\alpha(x_1,x_2,x_3) w_{x_2},\quad w_{x_1}=\beta(x_1,x_2,x_3) w_{x_3} .
\end{equation}
The system is compatible if and only if
\begin{equation}
  \alpha^2{\beta}_{x_2} -\beta^2\alpha_{x_3}+\beta\alpha_{x_1}-\alpha\beta_{x_1}=0.
  \label{compatibility}
\end{equation}
{\bf Proof}
~The compatibility condition is 
\begin{equation}
  (\partial_{x_1}-\alpha \partial_{x_2})(\partial_{x_1}-\beta \partial_{x_3})w-(\partial_{x_1}-\beta  \partial_{x_3})(\partial_{x_1}-\alpha \partial_{x_2})w=0.
\end{equation}
It is equivalent to:
\begin{equation}
  \left(\frac{\alpha \beta_{x_2}-\beta_{x_1}}{\beta}-\frac{\beta \alpha_{x_3}-\alpha_{x_1}}{\alpha}\right)u_{x_1}=0.
\end{equation}
The last expression is equivalent to \eqref{compatibility}.

In our case the governing equations in the solitary wave limit are
\begin{equation}
  \tau_t-u_q=0,\quad u_t-g/\tau^3 \tau_q=0,\quad F(n,\tau)_t+G(n,\tau)_q=0.     
  \label{solitary_limit_app} 
\end{equation}
Here, we replaced $h_1^{-1}$ by $\tau$ in the functions $F(n,h_1)$ and
$G(n, h_1)$ and used the same letters $F$ and $G$ for the functions of
new arguments.  Let us suppose that there exists the Riemann invariant
$R(n,\tau,u)$ such that
\begin{equation}
  R_t+\lambda R_q=0, \quad \lambda=\frac{G_n}{F_n}. 
\end{equation}
Expanding the equation for $R$, one obtains
\begin{equation}
  R_n (n_t+\lambda n_q)+R_\tau(\tau_t+\lambda \tau_q)+R_u (u_t+\lambda u_q)=0. 
\end{equation}
Also, replacing the time derivatives $n_t$, $u_t$ and $\tau_t$ from
\eqref{solitary_limit_app}, we obtain two equations corresponding to
the vanishing coefficients in front of the derivatives $u_q$ and
$\tau_q$. They are of the form
\begin{equation}
  R_n=\alpha(n,\tau)R_\tau, \quad R_n=\beta(n,\tau)R_u,
\end{equation}
with coefficients $\alpha$ and $\beta$ that are independent of $u$. We
take thus $x_1=n$, $x_2=\tau$ and $x_3=u$.  According to Theorem 1, the
compatibility condition is:
\begin{equation}
  \beta_\tau=\left(\frac{\beta}{\alpha}\right)_n.
\end{equation}
One can check with Mathematica software that this condition is not satisfied. Thus, the Riemann invariant coresponding to the eigenvalue $\lambda$ doesn't exist. 

\section{Numerical method}
         \label{sec:SGN_meth}

Let $p$ be the integrated fluid pressure divided by the
constant density $\rho$ and defined by
\[
  p= \frac{1}{2} h^{2}+
   \frac{h^{2}}{3}\frac{d^{2}h}{dt^{2}}.
\]
To solve the one-dimensional, homogeneous SGN
equations~\eqref{eq:186a}, \eqref{eq:186b} numerically, as
in~\cite{gavrilyuk_2d_2024} for the multidimensional SGN equations
over topography, we use the $\varpi$ formulation of the equations:
\begin{subequations}
  \label{eq:sgn-pi}
  \begin{align}
    &  h_{t}+
      \left (  hu \right )_{x} = 0,
      \label{eq:sgn-pi-mass} \\
    &  \left( hu \right)_{t} +
      \left ( hu^{2}+
      \frac{1}{2} h^{2} \right )_x = -\varpi_{x}.
      \label{eq:sgn-pi-mom}
  \end{align}
  Here $\displaystyle{\varpi= p-\frac{1}{2} h^{2}}$ denotes the
  averaged non-hydrostatic part of the pressure that is obtained by
  solving the linear elliptic problem
  \begin{align}
    \label{eq:sgn-pi-ellp}
    -\frac{h^{3}}{3}
    \left ( \frac{\varpi_{x}}{h} \right )_{x}+
    \varpi
    =  \frac{2}{3}
    h^{3} u_{x}^{2}+
    \frac{h^{3}}{3} h_{xx}.
  \end{align}
\end{subequations}
In the algorithm, we employ the hyperbolic-elliptic splitting approach
developed previously
in~\cite{le_metayer_numerical_2010,gavrilyuk_stationary_2020,gavrilyuk_2d_2024}.
This algorithm consists of two steps. In the first, hyperbolic step,
we employ a state-of-the-art method for the numerical solution of the
hyperbolic systems of equations~\eqref{eq:sgn-pi-mass},
\eqref{eq:sgn-pi-mom} over the time step $\Delta t$.  In the second,
elliptic step, using the approximate solution $h$ and $u$ computed
during the hyperbolic step, we numerically invert the elliptic
operator~\eqref{eq:sgn-pi-ellp} for $\varpi$ with prescribed boundary
conditions.

Note that in the hyperbolic step, rather than writing the equations in
the conservation form
$\boldsymbol{q}_{t} + \boldsymbol{f}(\boldsymbol{q})_{x} = 0$ with
$\boldsymbol{q} = (h, hu)^{T}$ and
$\boldsymbol{f} = \left (hu, hu^{2}+ \frac{1}{2} h^{2}+ \varpi
\right)^{T}$, which is essential in the conservative first-order
setting \cite{leveque_finite_2002} but is difficult to achieve more
than first order accuracy, we write it in the form of a balance law,
see~\cite{gavrilyuk_stationary_2020,gavrilyuk_2d_2024} for the details
of the numerical implementation.

\bibliographystyle{plain}
\bibliography{references}

\begin{thebibliography}{10}

\bibitem{olver_nist_2024}
{{NIST Digital Library}} of {{Mathematical Functions}}, 2024.

\bibitem{ablowitz_solitons_2018}
M.~J. Ablowitz, X.-D. Luo, and J.~T. Cole.
\newblock Solitons, the {{Korteweg-de Vries}} equation with step boundary
  values, and pseudo-embedded eigenvalues.
\newblock {\em J. Math. Phys.}, 59(9):091406, 2018.

\bibitem{ablowitz_solitonmean_2023}
Mark~J. Ablowitz, Justin~T. Cole, Gennady~A. El, Mark~A. Hoefer, and Xu-Dan
  Luo.
\newblock Soliton--mean field interaction in {{Korteweg}}--de {{Vries}}
  dispersive hydrodynamics.
\newblock {\em Stud. Appl. Math.}, 151(3):795--856, 2023.

\bibitem{benjamin_model_1972}
T.~B. Benjamin, J.~L. Bona, and J.~J. Mahony.
\newblock Model {{Equations}} for {{Long Waves}} in {{Nonlinear Dispersive
  Systems}}.
\newblock {\em Philosophical Transactions of the Royal Society of London.
  Series A, Mathematical and Physical Sciences}, 272(1220):47--78, 1972.

\bibitem{benzoni-gavage_modulated_2021}
Sylvie {Benzoni-Gavage}, Colin Mietka, and L.~Miguel Rodrigues.
\newblock Modulated equations of {{Hamiltonian PDEs}} and dispersive shocks.
\newblock {\em Nonlinearity}, 34(1):578, 2021.

\bibitem{biondini_soliton_2018}
Gino Biondini, Sitai Li, and Dionyssios Mantzavinos.
\newblock Soliton trapping, transmission, and wake in modulationally unstable
  media.
\newblock {\em Phys. Rev. E}, 98(4), 2018.

\bibitem{buhler_waves_2014}
Oliver B{\"u}hler.
\newblock {\em Waves and Mean Flows}.
\newblock Cambridge {{Monographs}} on {{Mechanics}}. Cambridge University
  Press, New York, NY, 2nd edition, 2014.

\bibitem{chew_waves_1999}
Weng~Cho Chew.
\newblock {\em Waves and {{Fields}} in {{Inhomogenous Media}}}.
\newblock John Wiley \& Sons, 1999.

\bibitem{congy_dispersive_2021}
T.~Congy, G.~A. El, M.~A. Hoefer, and M.~Shearer.
\newblock Dispersive {{Riemann}} problems for the
  {{Benjamin}}--{{Bona}}--{{Mahony}} equation.
\newblock {\em Stud. Appl. Math.}, 147(3):1089--1145, 2021.

\bibitem{el_resolution_2005}
G.~A. El.
\newblock Resolution of a shock in hyperbolic systems modified by weak
  dispersion.
\newblock {\em Chaos}, 15:037103, 2005.

\bibitem{el_theory_2007}
G.~A. El, A.~Gammal, E.~G. Khamis, R.~A. Kraenkel, and A.~M. Kamchatnov.
\newblock Theory of optical dispersive shock waves in photorefractive media.
\newblock {\em Phys. Rev. A}, 76(5):053813, 2007.

\bibitem{el_unsteady_2006}
G.~A. El, R.~H.~J. Grimshaw, and N.~F. Smyth.
\newblock Unsteady undular bores in fully nonlinear shallow-water theory.
\newblock {\em Phys. Fluids}, 18(2):027104--17, 2006.

\bibitem{el_asymptotic_2008}
G.~A. El, R.~H.~J. Grimshaw, and N.F. Smyth.
\newblock Asymptotic description of solitary wave trains in fully nonlinear
  shallow-water theory.
\newblock {\em Phys. D}, 237(19):2423--2435, 2008.

\bibitem{el_dispersive_2016-1}
G.~A. El and M.~A. Hoefer.
\newblock Dispersive shock waves and modulation theory.
\newblock {\em Physica D: Nonlinear Phenomena}, 333:11--65, 2016.

\bibitem{el_undular_2005}
G.~A. El, V.~V. Khodorovskii, and A.~V. Tyurina.
\newblock Undular bore transition in bi-directional conservative wave dynamics.
\newblock {\em Phys. D}, 206(3-4):232--251, 2005.

\bibitem{esler_dispersive_2011}
J.~G. Esler and J.~D. Pearce.
\newblock Dispersive dam-break and lock-exchange flows in a two-layer fluid.
\newblock {\em J. Fluid Mech.}, 667:555--585, 2011.

\bibitem{favrie_rapid_2017}
N.~Favrie and S.~Gavrilyuk.
\newblock A rapid numerical method for solving {{Serre}}--{{Green}}--{{Naghdi}}
  equations describing long free surface gravity waves.
\newblock {\em Nonlinearity}, 30(7):2718, 2017.

\bibitem{flaschka_multiphase_1980}
H.~Flaschka, M.~G. Forest, and D.~W. McLaughlin.
\newblock Multiphase averaging and the inverse spectral solution of the
  {{Korteweg-de Vries}} equation.
\newblock {\em Comm. Pure Appl. Math.}, 33:739--784, 1980.

\bibitem{gavrilyuk_large_1994}
S.~L. Gavrilyuk.
\newblock Large amplitude oscillations and their "thermodynamics" for continua
  with "memory".
\newblock {\em Eur J Mech BFluids}, 13(6):753--764, 1994.

\bibitem{gavrilyuk_kinematic_2015}
Sergey Gavrilyuk, Henrik Kalisch, and Zahra Khorsand.
\newblock A kinematic conservation law in free surface flow.
\newblock {\em Nonlinearity}, 28(6):1805--1821, 2015.

\bibitem{gavrilyuk_numerical_2024}
Sergey Gavrilyuk and Christian Klein.
\newblock Numerical study of the {{Serre}}--{{Green}}--{{Naghdi}} equations in
  {{2D}}.
\newblock {\em Nonlinearity}, 37(4):045014.

\bibitem{gavrilyuk_stationary_2020}
Sergey Gavrilyuk, Boniface Nkonga, Keh-Ming Shyue, and Lev Truskinovsky.
\newblock Stationary shock-like transition fronts in dispersive systems.
\newblock {\em Nonlinearity}, 33(10):5477--5509, 2020.

\bibitem{gavrilyuk_2d_2024}
Sergey Gavrilyuk and Keh-Ming Shyue.
\newblock {{2D Serre-Green-Naghdi Equations}} over {{Topography}}: {{Elliptic
  Operator Inversion Method}}.
\newblock {\em J. Hydraul. Eng.}, 150(1):04023054.

\bibitem{gavrilyuk_singular_2021}
Sergey Gavrilyuk and Keh-Ming Shyue.
\newblock Singular solutions of the {{BBM}} equation: Analytical and numerical
  study.
\newblock {\em Nonlinearity}, 35(1):388--410, 2021.

\bibitem{gavrilyuk_model_1995}
Sergey~L. Gavrilyuk and Denis Serre.
\newblock A {{Model}} of a {{Plug-Chain System Near}} the {{Thermodynamic
  Critical Point}}: {{Connection}} with the {{Korteweg Theory}} of
  {{Capillarity}} and {{Modulation Equations}}.
\newblock In Shigeki Morioka and Leen Van~Wijngaarden, editors, {\em {{IUTAM
  Symp}}. {{Waves Liq}}. {{Liq}}. {{Two-Phase Syst}}.}, pages 419--428.
  Springer Netherlands.

\bibitem{gavrilyuk_generalized_2001}
S.L. Gavrilyuk and V.M. Teshukov.
\newblock Generalized vorticity for bubbly liquidand dispersive shallow water
  equations.
\newblock {\em Contin. Mech. Thermodyn.}, 13(6):365--382, 2001.

\bibitem{green_derivation_1976}
A.E. Green and P.M. Naghdi.
\newblock A derivation of equations for wave propagation in water of variable
  depth.
\newblock {\em J. Fluid Mech.}, 78:237--246, 1976.

\bibitem{green_theory_1974}
Albert~Edward Green, N.~Laws, and P.~M. Naghdi.
\newblock On the theory of water waves.
\newblock {\em Proc. R. Soc. Lond. Math. Phys. Sci.}, 338(1612):43--55.

\bibitem{gurevich_nonlinear_1990}
A.~V. Gurevich, A.~L. Krylov, and G.~A. El.
\newblock Nonlinear modulated waves in dispersive hydrodynamics.
\newblock {\em Sov. Phys. JETP}, 71(5):899--910, 1990.

\bibitem{gurevich_nonstationary_1974}
A.~V. Gurevich and L.~P. Pitaevskii.
\newblock Nonstationary structure of a collisionless shock wave.
\newblock {\em Sov. Phys. JETP}, 38(2):291--297, 1974.

\bibitem{hoefer_shock_2014}
M.~A. Hoefer.
\newblock Shock {{Waves}} in {{Dispersive Eulerian Fluids}}.
\newblock {\em J. Nonlinear Sci.}, 24(3):525--577, 2014.

\bibitem{ivanov_motion_2022}
S.~K. Ivanov and A.~M. Kamchatnov.
\newblock Motion of dark solitons in a non-uniform flow of
  {{Bose}}--{{Einstein}} condensate.
\newblock {\em Chaos}, 32(11):113142, 2022.

\bibitem{jamshidi_long-wave_2020}
S.~Jamshidi and E.~R. Johnson.
\newblock The long-wave potential-vorticity dynamics of coastal fronts.
\newblock {\em J. Fluid Mech.}, 888:A19, 2020.

\bibitem{a_m_kamchatnov_nonlinear_2000}
A.~M. Kamchatnov.
\newblock {\em Nonlinear {{Periodic Waves}} and {{Their Modulations}}}.
\newblock World Scientific Publishing Company, 1st edition edition, 2000.

\bibitem{kamchatnov_hamiltonian_2024}
A.~M. Kamchatnov.
\newblock Hamiltonian theory of motion of dark solitons in the theory of
  nonlinear {{Schr{\"o}dinger}} equation.
\newblock {\em Theor Math Phys}, 219(1):567--575, 2024.

\bibitem{kamchatnov_propagation_2023}
A.~M. Kamchatnov and D.~V. Shaykin.
\newblock Propagation of generalized {{Korteweg--de Vries}} solitons along
  large-scale waves.
\newblock {\em Phys. Rev. E}, 108(5):054205, 2023.

\bibitem{lannes_water_2013}
David Lannes.
\newblock {\em The Water Waves Problem}, volume no. 188. of {\em Mathematical
  Surveys and Monographs}.
\newblock American Mathematical Society, Providence, RI, 2013.

\bibitem{le_metayer_numerical_2010}
O.~Le~M{\'e}tayer, S.~Gavrilyuk, and S.~Hank.
\newblock A numerical scheme for the {{Green}}--{{Naghdi}} model.
\newblock {\em J. Comput. Phys.}, 229(6):2034--2045, 2010.

\bibitem{leveque_finite_2002}
R.~J. LeVeque.
\newblock {\em Finite Volume Methods for Hyperbolic Problems}.
\newblock Cambridge University Press, 2002.

\bibitem{li_high_2014}
Maojun Li, Philippe Guyenne, Fengyan Li, and Liwei Xu.
\newblock High order well-balanced {{CDG}}--{{FE}} methods for shallow water
  waves by a {{Green}}--{{Naghdi}} model.
\newblock {\em Journal of Computational Physics}, 257:169--192.

\bibitem{li_hamiltonian_2002}
Yi~A Li.
\newblock Hamiltonian {{Structure}} and {{Linear Stability}} of {{Solitary
  Waves}} of the {{Green-Naghdi Equations}}.
\newblock {\em J. Nonlinear Math. Phys.}, 9:99--105.

\bibitem{lowman_dispersive_2013-1}
Nicholas~K. Lowman and M.~A. Hoefer.
\newblock Dispersive shock waves in viscously deformable media.
\newblock {\em J. Fluid Mech.}, 718:524--557, 2013.

\bibitem{maiden_solitary_2020}
M.~D. Maiden, N.~A. Franco, E.~G. Webb, G.~A. El, and M.~A. Hoefer.
\newblock Solitary wave fission of a large disturbance in a viscous fluid
  conduit.
\newblock {\em J. Fluid Mech.}, 883:A10, 2020.

\bibitem{maiden_solitonic_2018}
Michelle~D. Maiden, Dalton~V. Anderson, Nevil~A. Franco, Gennady~A. El, and
  Mark~A. Hoefer.
\newblock Solitonic {{Dispersive Hydrodynamics}}: {{Theory}} and
  {{Observation}}.
\newblock {\em Phys. Rev. Lett.}, 120(14):144101, 2018.

\bibitem{makarenko_second_1986}
N.~Makarenko.
\newblock A second long-wave approximation in the {{Cauchy-Poisson}} problem.
\newblock {\em Dyn. Cont. Media}, 77:56--72.

\bibitem{mao_observation_2023}
Yifeng Mao, Sathyanarayanan Chandramouli, Wenqian Xu, and Mark~A. Hoefer.
\newblock Observation of {{Traveling Breathers}} and {{Their Scattering}} in a
  {{Two-Fluid System}}.
\newblock {\em Phys. Rev. Lett.}, 131(14):147201, 2023.

\bibitem{mucalica_solitons_2022-1}
Ana Mucalica and Dmitry~E. Pelinovsky.
\newblock Solitons on the rarefaction wave background via the {{Darboux}}
  transformation.
\newblock {\em Proc. R. Soc. A.}, 478(2267):20220474, 2022.

\bibitem{olver_hamiltonian_1980}
Peter~J. Olver.
\newblock On the {{Hamiltonian}} structure of evolution equations.
\newblock {\em Math. Proc. Camb. Phil. Soc.}, 88(1):71--88, 1980.

\bibitem{ryskamp_oblique_2021}
S.~Ryskamp, M.~A. Hoefer, and G.~Biondini.
\newblock Oblique interactions between solitons and mean flows in the
  {{Kadomtsev}}--{{Petviashvili}} equation.
\newblock {\em Nonlinearity}, 34(6):3583--3617, 2021.

\bibitem{serre_contribution_1953}
F~Serre.
\newblock Contribution {\'a} l' {\'e}tude des {\'e}coulements permanents et
  variables dans les canaux.
\newblock {\em Houille Blanche}, (3, 6):374----388, 830----872, June--July,
  December, 1953.

\bibitem{sprenger_hydrodynamic_2018}
P.~Sprenger, M.~A. Hoefer, and G.~A. El.
\newblock Hydrodynamic optical soliton tunneling.
\newblock {\em Phys. Rev. E}, 97(3):032218, 2018.

\bibitem{su_korteweg-vries_1969}
C.H. Su and C.~S. Gardner.
\newblock Korteweg-de {{Vries}} equation and generalizations. {{III}}.
  {{Derivation}} of the {{Korteweg-de Vries}} and {{Burgers}} equation.
\newblock {\em J. Math. Phys.}, 10:536 -- 539, 1969.

\bibitem{suret_soliton_2024}
Pierre Suret, Stephane Randoux, Andrey Gelash, Dmitry Agafontsev, Benjamin
  Doyon, and Gennady El.
\newblock Soliton gas: {{Theory}}, numerics, and experiments.
\newblock {\em Phys. Rev. E}, 109(6):061001, 2024.

\bibitem{tkachenko_hyperbolicity_2020}
Sergey Tkachenko, Sergey Gavrilyuk, and Keh-Ming Shyue.
\newblock Hyperbolicity of the {{Modulation Equations}} for the
  {{Serre}}--{{Green}}--{{Naghdi Model}}.
\newblock {\em Water Waves}, 2(2):299--326, 2020.

\bibitem{sande_dynamic_2021}
Kiera van~der Sande, Gennady~A. El, and Mark~A. Hoefer.
\newblock Dynamic soliton--mean flow interaction with non-convex flux.
\newblock {\em J. Fluid Mech.}, 928:A21, 2021.

\bibitem{whitham_non-linear_1965}
G.~B. Whitham.
\newblock Non-linear dispersive waves.
\newblock {\em Proc. Roy. Soc. Ser. A}, 283:238--261, 1965.

\bibitem{whitham_linear_1999}
G.~B. Whitham.
\newblock {\em Linear and Nonlinear Waves}.
\newblock Wiley, New York, 1999.

\end{thebibliography}

\end{document}